\documentclass[aps,11pt,prd]{revtex4-1}
\usepackage{feynmp-auto}
\usepackage{amsmath}
\usepackage[utf8]{inputenc}
\usepackage{array}
\usepackage{multirow}
\usepackage{booktabs}
\usepackage{mathrsfs}
\usepackage[euler]{textgreek}
\usepackage{amsmath}
\usepackage{amsthm}
\usepackage{amsfonts}
\usepackage{amssymb}
\usepackage{graphicx}
\graphicspath{ {Figures/} }
\usepackage[font={small}]{caption}
\usepackage{subcaption}
\usepackage{float}
\usepackage[export]{adjustbox}
\usepackage[hang, flushmargin,bottom]{footmisc} 
\usepackage{hyperref}
\hypersetup{
     colorlinks   = true,
     citecolor    = blue
}
\usepackage{placeins}
\usepackage{enumitem}
\usepackage{slashed}
\usepackage{bbm}
\usepackage{appendix}

\usepackage{titlesec}
\titleformat{\chapter}[display]
  {\normalfont\LARGE\bfseries}
  {\chaptertitlename\ \thechapter}{5pt}{\LARGE}
  \titlespacing*{\chapter}{0pt}{-20pt}{35pt}
\usepackage{bigstrut}
\usepackage{pst-node}
\usepackage{pstricks}
\usepackage{physics}
\usepackage{mathtools}
\usepackage{setspace}
\usepackage{fancyhdr}
\usepackage[makeroom]{cancel}
\usepackage{feyn}
\newcommand{\tb}[1]{\textbf{#1}}
\newcommand{\be}{\begin{equation}}
\newcommand{\ee}{\end{equation}}
\newcommand{\bes}{\begin{equation*}}
\newcommand{\ees}{\end{equation*}}
\newcommand{\f}[2]{\frac{#1}{#2}}

\usepackage{hyperref}
\hypersetup{%
  colorlinks = true,
  linkcolor  = black
}
\usepackage{soul}

\begin{document}

\title{{\it{\bf{Leptophobic Dark Matter and the Baryon Number Violation Scale}}}}
\author{Pavel Fileviez P\'erez$^{1}$, Elliot Golias$^{1}$, Rui-Hao Li$^{1}$, Clara Murgui$^{2}$}
\affiliation{$^{1}$Physics Department and Center for Education and Research in Cosmology and Astrophysics (CERCA), 
Case Western Reserve University, Rockefeller Bldg. 2076 Adelbert Rd. Cleveland, OH 44106, USA \\
$^{2}$Departamento de F\'isica Te\'orica, IFIC, Universitat de Valencia-CSIC, 
E-46071, Valencia, Spain}
\email{pxf112@case.edu,elliot.golias@case.edu,ruihao.li@case.edu,clara.murgui@ific.uv.es}
\date{\today}
\begin{abstract}
We discuss the possible connection between the scale for baryon number violation and the cosmological bound on the dark matter relic density.
A simple gauge theory for baryon number which predicts the existence of a leptophobic cold dark matter particle candidate is investigated. 
In this context, the dark matter candidate is a Dirac fermion with mass defined by the new symmetry breaking scale. Using the 
cosmological bounds on the dark matter relic density we find the upper bound on the symmetry breaking scale around 200 TeV. 
The properties of the leptophobic dark matter candidate are investigated in great detail and we show the prospects to test 
this theory at current and future experiments. We discuss the main implications for the mechanisms to explain the matter
and antimatter asymmetry in the Universe.
\end{abstract}
\maketitle 



\section{INTRODUCTION}
%
The possible existence of dark matter in the Universe has called the attention of the scientific community for a long time. 
Fortunately, today we have many different types of experiments looking for possible signatures which can help us to reveal the nature of the dark 
matter~\cite{Bertone:2004pz}. There is a large list of candidates which can describe the properties of dark matter, but the so-called weakly interacting massive particles 
(WIMPs) are perhaps the most appealing candidates for the following reasons: a) One can predict the existence of stable or long-lived WIMPs 
in a large class of theories for physics beyond the Standard Model, b) we can compute, in a simple way, their relic density, c) since 
the relevant scale of new physics, in this case, is in the TeV range, one could expect missing energy signatures at the Large Hadron Collider, 
d) one could observe signatures in the different experiments looking for the recoil energy from WIMPs-nuclei (or WIMPs-electron) scatterings, 
or from the WIMPs annihilation products. This list of possibilities makes a strong case for WIMPs and motivates the different experiments to keep looking 
for  their signatures. Clearly, the direct confirmation of dark matter in these experiments would be one of the most spectacular discoveries in particle 
physics and cosmology.   

In the Standard Model of Particle Physics, the so-called baryon number (B) is an accidental global symmetry at the classical level which 
is broken at the quantum level by the $SU(2)$ instantons. In theories for physics beyond the Standard Model, we typically think about two 
main possibilities for baryon number violation: 1) Explicit breaking, and 2) Spontaneous Violation. The baryon number is explicitly broken in theories such 
as the Minimal Supersymmetric Standard Model, where one can have the so-called R-parity violating terms, or in Grand Unified Theories, where we have the unification 
of quarks and leptons, and the symmetry is broken at the very high scale $M_{GUT} \geq 10^{15}$ GeV. The only way to study the spontaneous 
breaking of baryon number is to think about theories where the baryon number is a local symmetry~
\cite{Pais:1973mi,Carone:1995pu,FileviezPerez:2010gw,FileviezPerez:2011pt,Duerr:2013dza,Perez:2014qfa}. These theories have been investigated recently in 
a series of papers and their main features are~\cite{Perez:2015rza}:

\begin{itemize}

\item One can define a simple anomaly free theory based on $U(1)_B$ which predicts the proton stability. In the context of these theories, there is no need to postulate 
the existence of a great desert between the electro-weak and high scales.

\item One predicts the existence of a cold dark matter candidate, and its mass is defined by the new symmetry scale.

\item The spontaneous breaking of baryon number at the low scale is possible in agreement with all experimental bounds in particle physics and cosmology.

\item A possible relation between the baryon asymmetry and dark matter densities is possible, and one can have a simple mechanism for baryogenesis in this 
context.

\end{itemize}

In this article, we investigate carefully the properties of the leptophobic dark matter candidate in a simple theory based on local baryon number.
In this theory, the dark matter candidate and the new leptophobic gauge boson masses are defined by the baryon number breaking scale. 
We study all dark matter annihilation channels in great detail and find that, in order to be in agreement with the cosmological bounds on the 
dark matter relic density, the local baryon symmetry must be broken below the ${\cal O}(10^2)$ TeV scale. This upper bound coming from cosmology has profound 
implications because it tells us that the simplest theories for spontaneous baryon number violation can be tested in the near future at collider experiments
and predict different signatures in dark matter experiments. 

This article is organized as follows: In section 2, we discuss a simple effective theory for leptophobic dark matter models, while in section 3, we discuss in detail a simple theory based on the local 
baryon number, and discuss some of the main experimental constraints. In section 4, we discuss in great detail the properties of the cold dark matter candidate; 
we discuss all possible annihilation channels and show the parameter space allowed by the relic density constraints, direct, and indirect detection bounds.

\section{EFFECTIVE THEORY FOR LEPTOPHOBIC DARK MATTER}
The theory we investigate in this article predicts a fermionic dark matter candidate. In this case the cold dark matter is a 
Standard Model singlet $\chi$ with very suppressed coupling to leptons, i.e. leptophobic.
One could imagine different theories which could predict the existence of leptophobic dark matter candidates.
In this section we discuss the different effective operators one could have in scenarios with a leptophobic dark matter candidate.
 
Integrating out heavy fields in simple extensions of the Standard Model we can obtain a simple effective field theory for the leptophobic 
dark matter. In the theories we are interested in, one can expect the following dimension five and six operators defining the interactions between the 
Standard Model fields and the dark matter field $\chi$:
\begin{eqnarray}
{\cal{O}}_1 &=& \frac{c_1}{\Lambda} \bar{\chi} (a_\chi + i b_\chi \gamma_5) \chi H^\dagger H,\\
{\cal{O}}_2 &=& \frac{c_2}{\Lambda^2} \bar{\chi} (A_\chi + B_\chi \gamma_5) \gamma^\mu \chi \bar{Q}_L (A_Q + B_Q \gamma_5) \gamma_\mu Q_L, \\
{\cal{O}}_3 &=&  \frac{c_3}{\Lambda^2} \bar{\chi} (A_\chi + B_\chi \gamma_5) \gamma^\mu \chi \bar{u}_R (A_u + B_u \gamma_5) \gamma_\mu u_R, \\
{\cal{O}}_4 &=&  \frac{c_4}{\Lambda^2} \bar{\chi} (A_\chi + B_\chi \gamma_5) \gamma^\mu \chi \bar{d}_R (A_d + B_d \gamma_5) \gamma_\mu d_R,  \\
{\cal{O}}_5 &=&  \frac{c_5}{\Lambda^2}( \bar{Q}_L  P_R \chi ) ( \bar{\chi} P_L Q_L), \\
{\cal{O}}_6 &=&  \frac{c_6}{\Lambda^2}( \bar{u}_R  P_L \chi ) ( \bar{\chi} P_R u_R), \\
{\cal{O}}_7 &=& \frac{c_7}{\Lambda^2}( \bar{d}_R  P_L \chi ) ( \bar{\chi} P_R d_R).
\end{eqnarray}   
Here, the multiplets $Q_L$, $u_R$, $d_R$, and $H$ are the Standard Model multiplets listed in Table 1. The simplest and most motivated models for leptophobic 
dark matter are based on $U(1)_B$, where $B$ is the baryon number. These models have been proposed 
in Refs.~\cite{Duerr:2013dza,Duerr:2013lka,Perez:2014qfa,Perez:2015rza}. In the models based on local baryon 
number, we get only a set of the operators listed above once we integrate out the new heavy degrees of freedom. We will examine the simplest theories in the next section and discuss 
the origin of these effective operators. The operator ${\cal{O}}_1$ is generated once we integrate out the Higgs breaking $U(1)_B$, but $b_\chi=0$; the operators ${\cal{O}}_2$, ${\cal{O}}_3$, 
and ${\cal{O}}_4$ are generated once we integrate out the new gauge boson associated to baryon number, but in this case $A_Q=B_Q=A_u=B_u=A_d=B_d=1/3$. 
The operators ${\cal{O}}_6$, ${\cal{O}}_7$, and ${\cal{O}}_8$ are not generated in the simplest models for $U(1)_B$ because one does not have colored scalar fields.
For recent studies in models with leptophobic dark matter candidates see Refs.~\cite{Ohmer:2015lxa,Duerr:2016tmh,Duerr:2014wra,Ellis:2018xal,ElHedri:2018cdm,Caron:2018yzp,Krovi:2018fdr,Gondolo:2011eq,Batell:2014yra}, and for a complete 
list of effective operators in dark matter models see, for example, Ref.~\cite{Bishara:2018vix}. 
In the next section, we will discuss the main features of the simplest model for the local baryon number and the properties of the dark matter candidate.
%

\section{THEORY FOR BARYON NUMBER}
The simplest realistic theories for the spontaneous breaking of baryon number have been proposed in Refs.~\cite{FileviezPerez:2011pt,Duerr:2013dza,Perez:2014qfa}.
These theories are based on the local gauge symmetry
\begin{equation}
SU(3)_C\otimes SU(2)_L\otimes U(1)_Y\otimes U(1)_B. \nonumber 
\end{equation}
Here, we will study the simplest theory for baryon number where the anomalies are cancelled with colorless fields~\cite{Duerr:2013dza}. In Table I, we list the particle content including the 
Standard Model content, the new fermionic fields needed for anomaly cancellation, and a new Higgs needed for the spontaneous breaking of baryon number. 
%
\begin{table}[h]\setlength{\bigstrutjot}{6pt}
\centering
\begin{tabular}{|ccccc|}\hline
    Fields & $SU(3)_C$ & $SU(2)_L$ & $U(1)_Y$  & $U(1)_B$ \bigstrut\\\hline\hline
    $\ell^i_L = \mqty(\nu_L^i\\ e^i_L)$ & $\tb{1}$ & $\tb{2}$ & $-\frac{1}{2}$ & $0$\bigstrut\\
    $e^i_R$ & $\tb 1$ & $\tb 1$ & $-1$ & $0$  \bigstrut\\
    $Q_L = \mqty(u_L^i\\ d^i_L)$ & $\tb 3$ & $\tb 2$ & $\frac{1}{6}$ & $\frac{1}{3}$ \bigstrut\\
    $u^i_R$ & $\tb 3$ & $\tb 1$ & $\frac{2}{3}$ & $\frac{1}{3}$ \bigstrut\\
    $d^i_R$ & $\tb 3$ & $\tb 1$ & $-\frac{1}{3}$ & $\frac{1}{3}$ \bigstrut\\
    $H$ & $\tb 1$ & $\tb 2$ & $\f{1}{2}$& 0 \bigstrut\\ \hline 
    $\Psi_L = \mqty(\Psi_L^0 \\ \Psi_L^-)$ & $\tb{1}$ & $\tb{2}$ & $Y_1$ & $B_1$\bigstrut\\
    $\Psi_R = \mqty(\Psi_R^0 \\ \Psi_R^-)$ & $\tb 1$ & $\tb 2$ & $Y_1$ & $B_2$  \bigstrut\\
    $\eta_R$ & $\tb 1$ & $\tb 1$ & $Y_2$ & $B_1$ \bigstrut\\
    $\eta_L$ & $\tb 1$ & $\tb 1$ & $Y_2$ & $B_2$ \bigstrut\\
    $\chi_R$ & $\tb 1$ & $\tb 1$ & $Y_3$ & $B_1$ \bigstrut\\
    $\chi_L$ & $\tb 1$ & $\tb 1$ & $Y_3$ & $B_2$ \bigstrut\\
     $S_B$ & $\tb 1$ & $\tb 1$ & $0$ & $-3$ \bigstrut\\ \hline
  \end{tabular}
  \caption{Particle Content, $i = 1, 2, 3$ is the family index.}
  \label{taba1}
\end{table}

The Lagrangian of the theory can be written as
\begin{eqnarray}
{\mathscr{L}}_{B} &=& {\cal{L}}_{SM} - \frac{g_B}{3} \left( \bar{Q}_L \gamma^\mu Z_\mu^B Q_L + \bar{u}_R \gamma^\mu Z_\mu^B u_R 
+ \bar{d}_R \gamma^\mu Z_\mu^B d_R \right) \nonumber \\
&+&  i\overline{\Psi}_L\slashed{D}\Psi_L + i\overline{\Psi}_R\slashed{D}\Psi_R + i\overline{\chi}_L\slashed{D}\chi_L 
+ i\overline{\chi}_R\slashed{D}\chi_R + 
i\overline{\eta}_L\slashed{D}\eta_L + i\overline{\eta}_R\slashed{D}\eta_R \nonumber \\
& + &  (D_\mu S_B)^\dagger (D_\mu S_B) - V(H, S_B)
- ( y_1\overline\Psi_L H\eta_R + y_2\overline\Psi_L \widetilde H\chi_R + y_3 \overline \Psi_R H\eta_L \nonumber \\
&+& y_4\overline\Psi_R\widetilde H\chi_L
 + \lambda_\Psi\overline\Psi_L\Psi_R S_B + \lambda_\eta\overline\eta_R\eta_L S_B + \lambda_\chi\overline \chi_R\chi_L S_B \ + \ \rm{h.c.} ),
\end{eqnarray} 
where  ${\cal{L}}_{SM}$ is the Lagrangian of the Standard Model, $\widetilde H = i\sigma_2 H^*$, and $V(H, S_B)$ contains all the relevant terms for the scalar fields excluding the Standard Model Higgs potential. 
Anomaly cancellation requires the following relation between the new hypercharges~\cite{Duerr:2013dza}:
\begin{equation}
Y_2^2 + Y_3^2 - 2 Y_1^2 =1/2.
\end{equation}
In our study, we will investigate the case $Y_3=0$, $Y_1=-1/2$ and $Y_2=-1$ for simplicity, and we will show that, in this context, we have a viable dark matter candidate. 
Here, we write the most generic interaction terms without assuming any particular relation between the baryon numbers $B_1$ and $B_2$. 
However, anomaly cancellation requires 
\begin{equation}
B_1 - B_2 = -3,
\end{equation} 
and then $S_B$ must have baryon number $-3$. The above relation is a key prediction of the theory; the proton is absolutely stable, and therefore, the symmetry can be broken at the low scale. 
For more details about these models see the review in Ref.~\cite{Perez:2015rza}.
%
\subsection{Spontaneous Symmetry Breaking}
%
The Higgs sector of the theory is composed of the Standard Model Higgs and the new Higgs $S_B$ breaking the local gauge symmetry. They are given by
\be
H = \mqty(h^+\\ \f{1}{\sqrt 2}\qty(h_0 + i A_0)), \ {\rm{and}} \  S_B = \f{1}{\sqrt 2}\qty(h_B + i A_B),
\ee
and the scalar potential reads as
\be
V=-\mu_H^2 H^\dagger H + \lambda_H \left( H^\dagger H \right)^2 -\mu_B^2 S_B^\dagger S_B + \lambda_B \left( S_B^\dagger S_B \right)^2 + \lambda_{HB} \left( H^\dagger H \right)\left( S_B^\dagger S_B \right).
\ee
Once $S_B$ acquires the vacuum expectation value $\ev{S_B}= \f{1}{\sqrt 2} v_B$, the local symmetry $U(1)_B$ is broken to a global symmetry $U(1)_\chi$. 
This global symmetry is anomaly free and acts non-trivially on the new fermionic fields: $$\Psi_L \to e^{i \chi} \Psi_L, \Psi_R \to e^{i \chi} \Psi_R, \eta_L \to e^{i \chi} \eta_L, 
\eta_R \to e^{i \chi} \eta_R, \chi_L \to e^{i \chi} \chi_L, \chi_R \to e^{i \chi} \chi_R.$$ 
Therefore, if the lightest new field in this sector is neutral, it can be stable 
and a good candidate for the cold dark matter in the Universe. We will study the properties of this dark matter candidate in detail. The rest of the symmetry is broken as in the Standard Model, and we have nothing to add. 
See Ref.~\cite{Duerr:2017whl} for a recent discussion of the Higgs sector of this type of models. We would like to mention that, after symmetry breaking, there are two global anomaly free symmetries: $B-L$ in the Standard Model sector and $U(1)_\chi$ in the new fermionic sector.

In this theory, there are two physical Higgses, $h_1$ and $h_2$, where $h_1$ corresponds to the Standard Model-like Higgs. They are defined as
\begin{eqnarray}
h_1 &=& h_0 \cos \theta_B + h_B \sin \theta_B, \\
h_2 &=& h_B \cos \theta_B - h_0 \sin \theta_B,
\end{eqnarray}
where the mixing angle $\theta_B$ is given by
\be
\tan 2 \theta_B = \frac{ \lambda_{HB} v_0 v_B}{ \lambda_H v_0^2 - \lambda_B v_B^2}.
\ee
The physical masses for the Higgses are given by
\begin{eqnarray}
M_{h_1}^2 &=& \lambda_H v_0^2 + \lambda_B v_B^2 - (\lambda_B v_B^2 - \lambda_H v_0^2 ) \left(1 + \frac{\lambda_{HB}^2 v_0^2 v_B^2 }{(\lambda_H v_0^2 - \lambda_B v_B^2 )^{2}} \right)^{1/2},\\
M_{h_2}^2 &=& \lambda_H v_0^2 + \lambda_B v_B^2 + (\lambda_B v_B^2 - \lambda_H v_0^2 ) \left(1 + \frac{\lambda_{HB}^2 v_0^2 v_B^2 }{(\lambda_H v_0^2 - \lambda_B v_B^2 )^{2}} \right)^{1/2}.
\end{eqnarray}

\begin{figure}[t]
\centering
\includegraphics[width=0.8\linewidth]{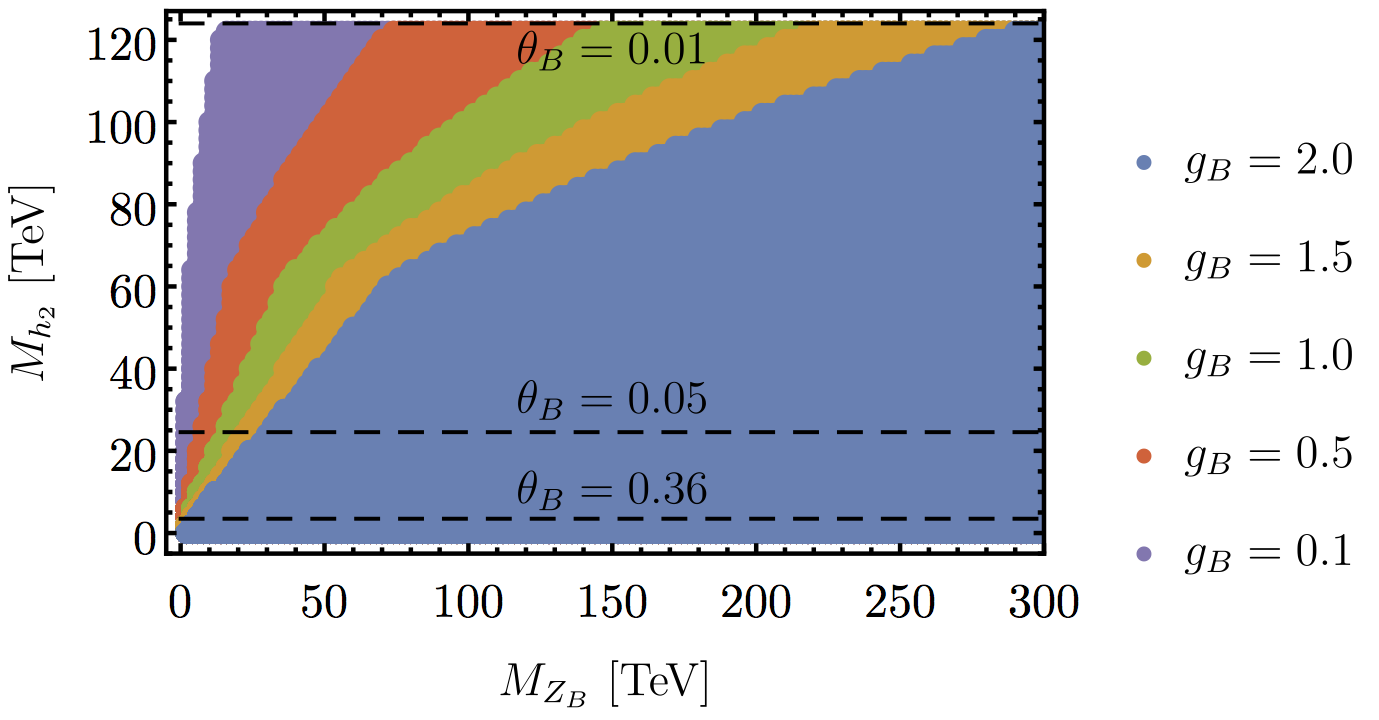} 
\caption{Parameter space in the $M_{h_2}$--$M_{Z_B}$ plane allowed by perturbativity bounds, $\lambda_H, \lambda_B, \lambda_{HB} \leq 4 \pi$, 
and the condition on the scalar potential (bounded from below). Here, the mixing angle $\theta_B$ changes from 0.01 to 0.36, and the different colors correspond to the different values for the gauge coupling $g_B$.}
\label{Mh2-values}
\end{figure}

The quartic couplings in the scalar potential can be written as a function of the Higgs masses and the mixing angle:
\begin{align}
 \lambda_H    &= \frac{1}{4 v_0^2} \left[ M_{h_1}^2 + M_{h_2}^2 + \left( M_{h_1}^2 - M_{h_2}^2\right) \cos 2 \theta_B \right], \\
 \lambda_B    &= \frac{1}{4 v_B^2} \left[ M_{h_1}^2 + M_{h_2}^2 + \left( M_{h_2}^2 - M_{h_1}^2\right) \cos 2 \theta_B \right] , \\
 \lambda_{HB} &= \frac{1}{2 v_0 v_B} \left( M_{h_1}^2 - M_{h_2}^2 \right) \sin 2 \theta_B.
\end{align}
We note that, in order to have a potential bounded from below, the following condition must be satisfied:
\begin{equation}
\lambda_H \lambda_B - \frac{1}{4} \lambda_{HB}^2  > 0,
\end{equation}
and the perturbativity condition imposes that $\lambda_H \leq 4 \pi$, $\lambda_B \leq 4 \pi$, and $\lambda_{HB} \leq 4 \pi$.
The mass of the new gauge boson is given by 
\begin{equation}
M_{Z_B}=3 g_B v_B,
\end{equation} 
and it couples only to quarks and the new fermions present in the theory, i.e. we have a leptophobic gauge boson in the theory.
In Fig.~\ref{Mh2-values}, we show the numerical values for the mass of the second Higgs in the $M_{h_2}-M_{Z_B}$ plane allowed by the perturbativity bounds, i.e. $\lambda_H, \lambda_B, \lambda_{HB} \leq 4 \pi$, and the condition on the scalar potential (bounded from below). 
Here, the mixing angle $\theta_B$ changes from 0.01 to 0.36, and the different colors correspond to the different values for the gauge coupling $g_B$.
We note that the maximal allowed value for the mixing value is around 0.36, see discussion in Ref.~\cite{Duerr:2017whl}.
One can clearly see that, for a large mixing angle, there is a very strong upper bound on the Higgs mass of the new Higgs, see Fig.~\ref{Mh2-values}. 
We will take into account all these results and conditions on the different parameters for our numerical studies in the next section.
%
\subsection{Leptophobic Gauge Boson}
%
\begin{figure}[h]
\begin{minipage}[t]{0.65\textwidth}
\includegraphics[width=0.9\linewidth]{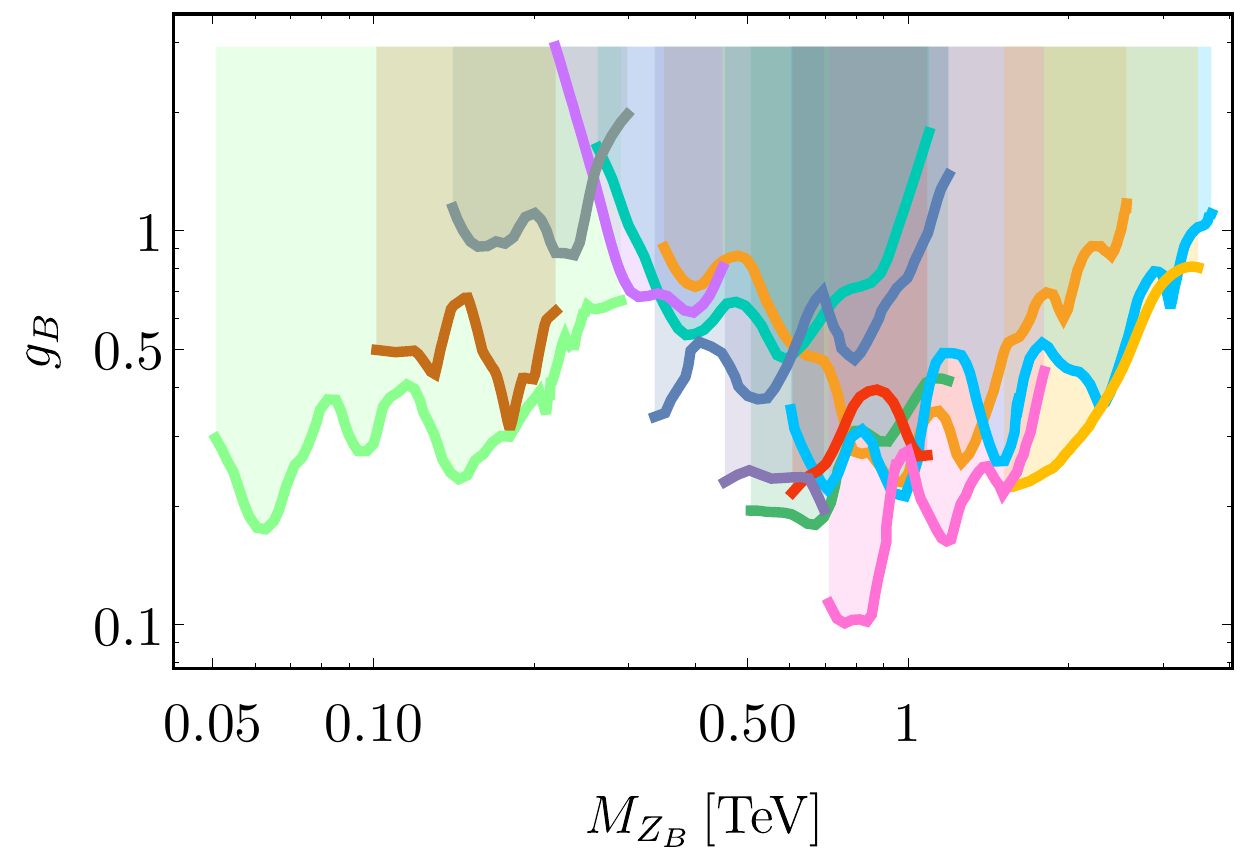}
\includegraphics[width=0.9\linewidth]{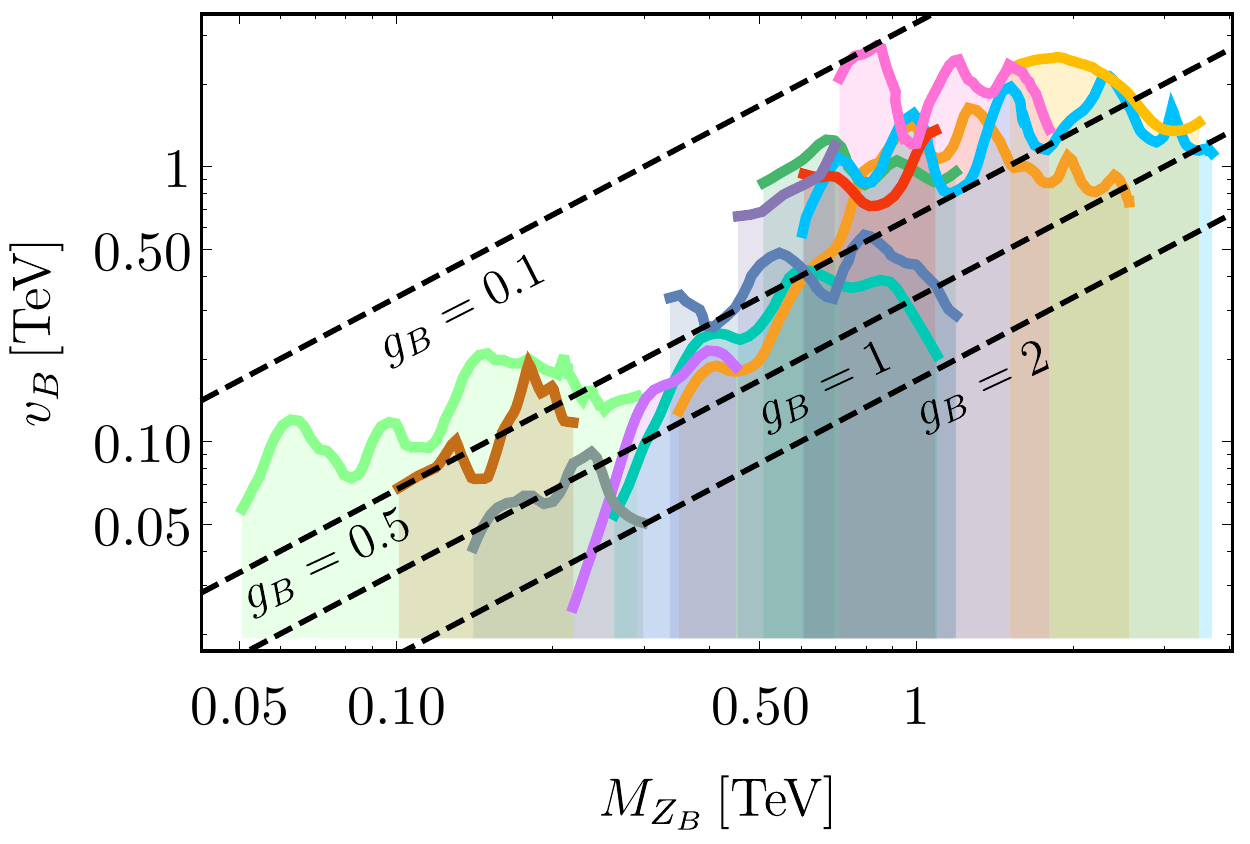}
\end{minipage}
\begin{minipage}[t]{.3\textwidth}
\includegraphics[width=0.9\linewidth]{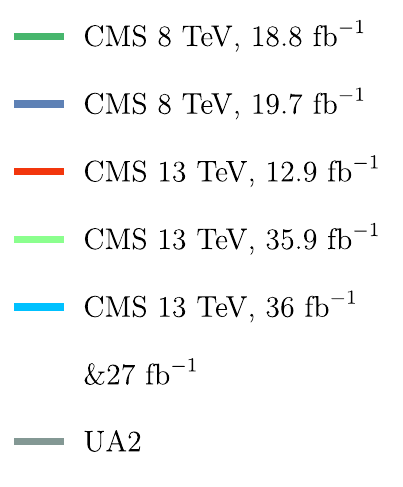}
\includegraphics[width=0.94\linewidth]{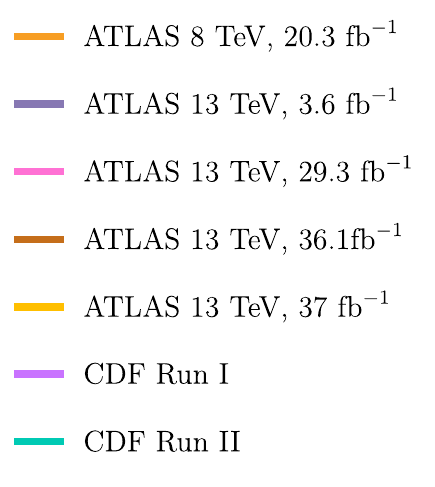}
\end{minipage}
\caption{Experimental bounds for the leptophobic gauge boson $Z_B$. Here, we use the CMS analyses (8 TeV and 18.8 $\text{fb}^{-1}$\cite{CMS18.8}, 8 TeV and 19.7 $\text{fb}^{-1}$\cite{CMS19.7}, 13 TeV and 12.9 $\text{fb}^{-1}$, 13 TeV and 35.9 $\text{fb}^{-1}$\cite{CMS35.9}, 13 TeV and 36 $\text{fb}^{1-}$ \& 27 $\text{fb}^{-1}$\cite{CMS27&36}), ATLAS results (8 TeV and 20.3 $\text{fb}^{-1}$ \cite{ATLAS20.3}, 13 TeV and 3.6 $\text{fb}^{-1}$ and 29.3 $\text{fb}^{-1}$\cite{ATLAS3.6&29.3}, 13 TeV and 36.1 $\text{fb}^{-1}$\cite{ATLAS36.1}, 13 TeV and 37 $\text{fb}^{-1}$\cite{ATLAS37.0}), and other experiments (UA2~\cite{UA2} and CDF~\cite{CDF}).} 
\label{ZBbounds}
\end{figure}
This theory predicts the existence of a leptophobic gauge boson $Z_B$, and it is important to know the experimental bounds 
on its mass and the gauge coupling $g_B$ associated with the baryon number. Therefore, in this way, we can infer what is the lower 
bound on the symmetry breaking scale. In Fig.~\ref{ZBbounds} (upper panel), we show the experimental bounds in the $g_B-M_{Z_B}$ plane. Concretely, we show 
the experimental bounds from CMS~\cite{CMS35.9, CMS27&36, CMS19.7, CMS18.8}, ATLAS~\cite{ATLAS36.1, ATLAS3.6&29.3, ATLAS37.0, ATLAS20.3}, 
UA2~\cite{UA2}, and CDF~\cite{CDF} experiments. As one can appreciate, the leptophobic gauge boson can be very light without assuming a very small coupling $g_B$. 
Using the relation $M_{Z_B}=3 g_B v_B$, we can make use of the same experimental results to show the lower bound on the symmetry breaking scale $v_B$, 
see Fig.~\ref{ZBbounds} (lower panel). It is striking to see that the local $U(1)_B$ can be spontaneously broken at the low scale in agreement with all collider bounds.
%
\subsection{New Fermion Masses}

In the theory discussed above, one predicts the existence of new neutral and charged fermions. After symmetry breaking, we can compute 
the mass matrix of the new neutral fermions in the basis $\left( \chi_L^0  \  (\chi^c)_L  \  \Psi_L^0  \  (\Psi^c)_L \right)$, and it is given by
\be
\mathscr{M}_0 = 
{\mqty(0 & M_\chi & 0 & M_4 \\ M_\chi & 0 & M_2 & 0 \\ 0 & M_2 & 0 & M_\Psi \\ M_4 & 0 & M_\Psi & 0)},
\ee
 where
\be
\qquad M_\Psi = \f{\lambda_\Psi v_B}{\sqrt 2},\qquad M_\chi =  \f{\lambda_\chi v_B}{\sqrt 2},\qquad M_2 = \f{y_2v_0}{\sqrt 2},\qquad  {\rm{and}} \qquad M_4 = \f{y_4v_0}{\sqrt 2}.
\ee
We can diagonalize the above mass matrix using 
\be
\mqty(X_1 \\ X_2\\ X_3\\ X_4)_L = U \mqty(\chi_L\\ (\chi^c)_L \\ \Psi_L \\ (\Psi^c)_L),
\ee
such that $U^T {\mathscr{M}_0} \ U = {\mathscr{M}}_0^{diag}$, where ${\mathscr{M}}_0^{diag}= {\rm{diag}} \left( M_{X_1}, M_{X_2}, M_{X_3}, M_{X_4} \right)$ contains all physical masses for the new neutral fermions. We note that there is no mixing between the Standard Model fermions and the new fermions.

The mass matrix for the new charged fermions is given by 
\be
\mathscr{M}_{\pm} = 
{\mqty(M_\Psi & M_1  \\ M_3 & M_\eta )},
\ee
in the basis $\left( \Psi_L^- \  \eta_L^- \right)$ and $\left( \Psi_R^- \ \eta_R^- \right)$. The different masses in the above mass matrix are given by

\be
M_\eta= \lambda_\eta v_B / \sqrt{2}, \qquad M_1 =  y_1 v_0 /  \sqrt{2},  \qquad  {\rm{and}} \qquad M_3 = y_3 v_0 / \sqrt{2}. 
\ee
In our convention, this mass matrix is diagonalized by $V_L^\dagger \mathscr{M}_{\pm}  V_R = \mathscr{M}_{\pm}^{diag}$, where $V_L$ 
and $V_R$ are defined by the following relations
\be
\mqty(X_1^- \\ X_2^-)_L = V_L \mqty(\Psi_L^- \\ \eta_L^-), \qquad  {\rm{and}} \qquad \mqty(X_1^- \\ X_2^-)_R = V_R \mqty(\Psi_R^- \\ \eta_R^-).
\ee

In this paper, we will investigate the dark matter properties in the limit when $y_2$ and $y_4$ are very small because, only in this case, we can avoid large 
interactions between our dark matter candidate and the $Z$ gauge boson. In Appendix B, we study the case where the dark matter candidate is a pure $SU(2)_L$ candidate; 
we can see, in Fig.~\ref{DDPsiPlot}, the predictions for the direct detection cross section mediated by the Standard Model $Z$ gauge boson. As one can appreciate, 
this case is excluded by the experiment. Hence, we focus on the scenario where the dark matter candidate is a Standard Model singlet.  
In this limit, our dark matter is a Dirac fermion $$\chi=\chi_L + \chi_R,$$ with mass $M_\chi$, defined by the scale of symmetry breaking. In the next section, we investigate in great detail the properties of this dark matter candidate.

\section{ LEPTOPHOBIC DARK MATTER}
%
\begin{figure}[t]
\centering
\includegraphics[width=0.9\linewidth]{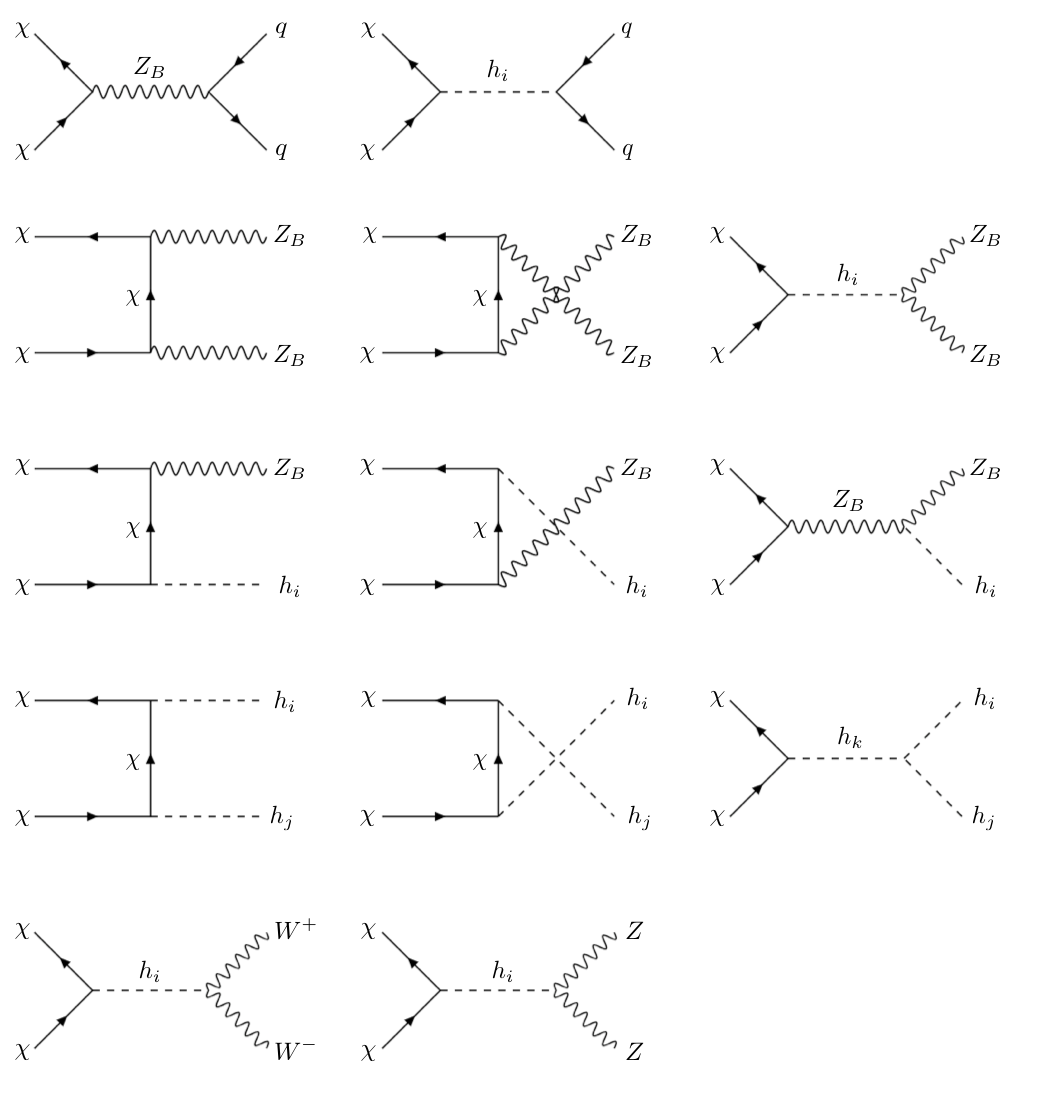} 
\caption{Dark Matter Annihilation Channels.}
\label{Annihilation}
\end{figure}

The lightest new fermion in the theory discussed above can be a good candidate for the cold dark matter if it is neutral. 
In the previous section, we discussed the properties of the new fermions present in the theory.  Since the direct detection bounds 
are very strong for any dark matter field with $SU(2)_L$ quantum numbers, we investigate the main and more general scenario when the dark matter is a Dirac fermion: 
$\chi=\chi_L + \chi_R$. 
%
\subsection{Relic Density}
In Fig.~\ref{Annihilation}, we show all the possible annihilation channels for our dark matter candidate $\chi$. This simple theory for dark matter 
has the following free parameters: $$g_B, B, M_\chi, M_{Z_B}, \theta_B, \ \text{and} \ M_{h_2},$$  where $B=B_1 + B_2$ is the total baryon number.
Knowing all annihilation channels, we can use the analytic approximation to compute the relic density~\cite{Gondolo:1990dk}
\begin{equation}
\label{eq:relicdensity}
\Omega_\text{DM} h^2 = \frac{1.05 \times 10^{9}{\rm{GeV}}^{-1}}{J(x_f) \ M_\text{Pl}},
\end{equation}
where $M_\text{Pl}=1.22 \times 10^{19}\ {\rm{GeV}}$ is the Planck scale, 
$g_\ast$ is the total number of effective relativistic degrees of freedom at the time of freeze-out, 
and the function $J(x_f)$ reads as
\begin{equation}
J(x_f)=\int_{x_f}^{\infty} \frac{g_\ast^{1/2}(x) \langle \sigma v \rangle (x)}{x^2} dx.
\end{equation}
The thermally averaged annihilation cross section times velocity $\langle \sigma v \rangle$ is a function of $x=M_\chi/T$ and is given by
\begin{equation}
 \langle\sigma v\rangle (x) = \frac{x}{8 M_\chi^5 K_2^2(x)} \int_{4 M_\chi^2}^\infty \sigma \times ( s - 4 M_\chi^2) \ \sqrt{s} \ K_1 \left(\frac{x \sqrt{s}}{M_\chi}\right) ds,
\end{equation}
where $K_1(x)$ and $K_2(x)$ are the modified Bessel functions.
The freeze-out parameter $x_f$ can be computed using
\begin{equation}
x_f= \ln \left( \frac{0.038 \ g \ M_\text{Pl} \ M_\chi \ \langle\sigma v\rangle (x_f) }{\sqrt{g_\ast x_f}} \right),
\end{equation}
where $g$ is the number of degrees of freedom of the dark matter particle. In order to discuss our numerical results, we will focus on two main scenarios which give a global perspective of the whole spectrum:

\begin{itemize}

\item Minimal Mixing Scenario

When there is no mixing between the two Higgses present in the theory $(\theta_B=0)$ , the main dark matter annihilation channels are:

$$\bar{\chi} \chi \to \bar{q} q, Z_B Z_B, Z_B h_2, h_2 h_2.$$  

In Fig.~\ref{Annihilation-Brs-minimal-mixing}, we show the different branching ratios for the channels mentioned above. 
For illustration, we use the following values for the input parameters:  $M_{Z_B}=3$ TeV, $M_{h_2}=1$ TeV, $g_B=0.5$, $x_f=24$, and $B=-1$.
As one can appreciate, for dark matter masses below and close to the resonance (i.e. $\lesssim $ 2 TeV), the dominant annihilation channel corresponds to the annihilation into two quarks, 
while for masses larger than the $Z_B$ boson mass (i.e. 3 TeV), the dominant annihilation channel is $\bar{\chi} \chi \to Z_B h_2$. 
\begin{figure}[h]
\centering
\includegraphics[width=0.8\linewidth]{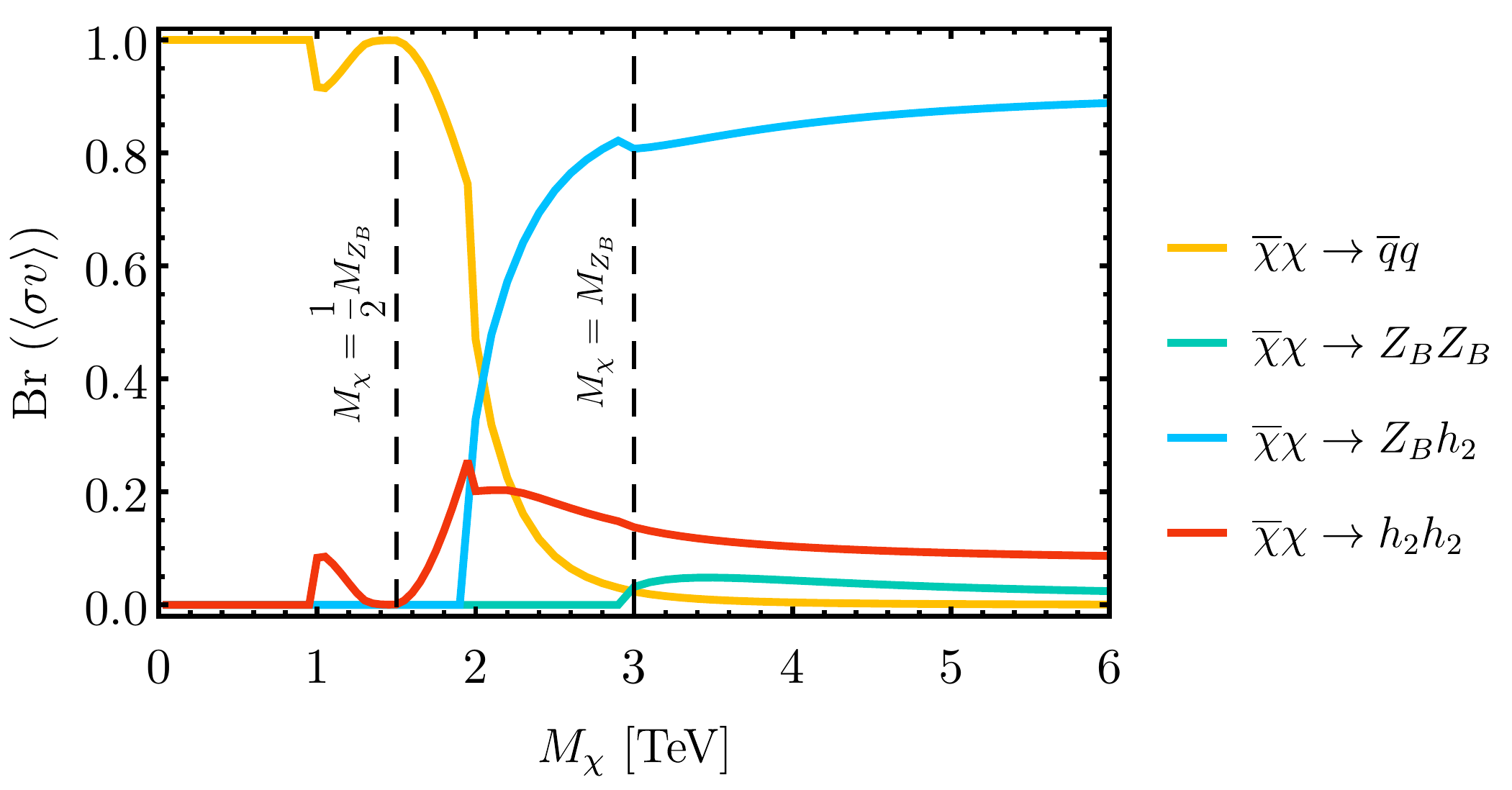} \\
\caption{Branching ratios for the different dark matter annihilation channels when the mixing angle between the Higgses is $\theta_B=0$. For illustration, we use the following values for the input parameters: 
$M_{Z_B}=3$ TeV, $M_{h_2}=1$ TeV, $g_B=0.5$, $x_f=24$, and $B=-1$. }
\label{Annihilation-Brs-minimal-mixing}
\end{figure}
\begin{figure}[h]
\centering
\includegraphics[width=0.48\linewidth]{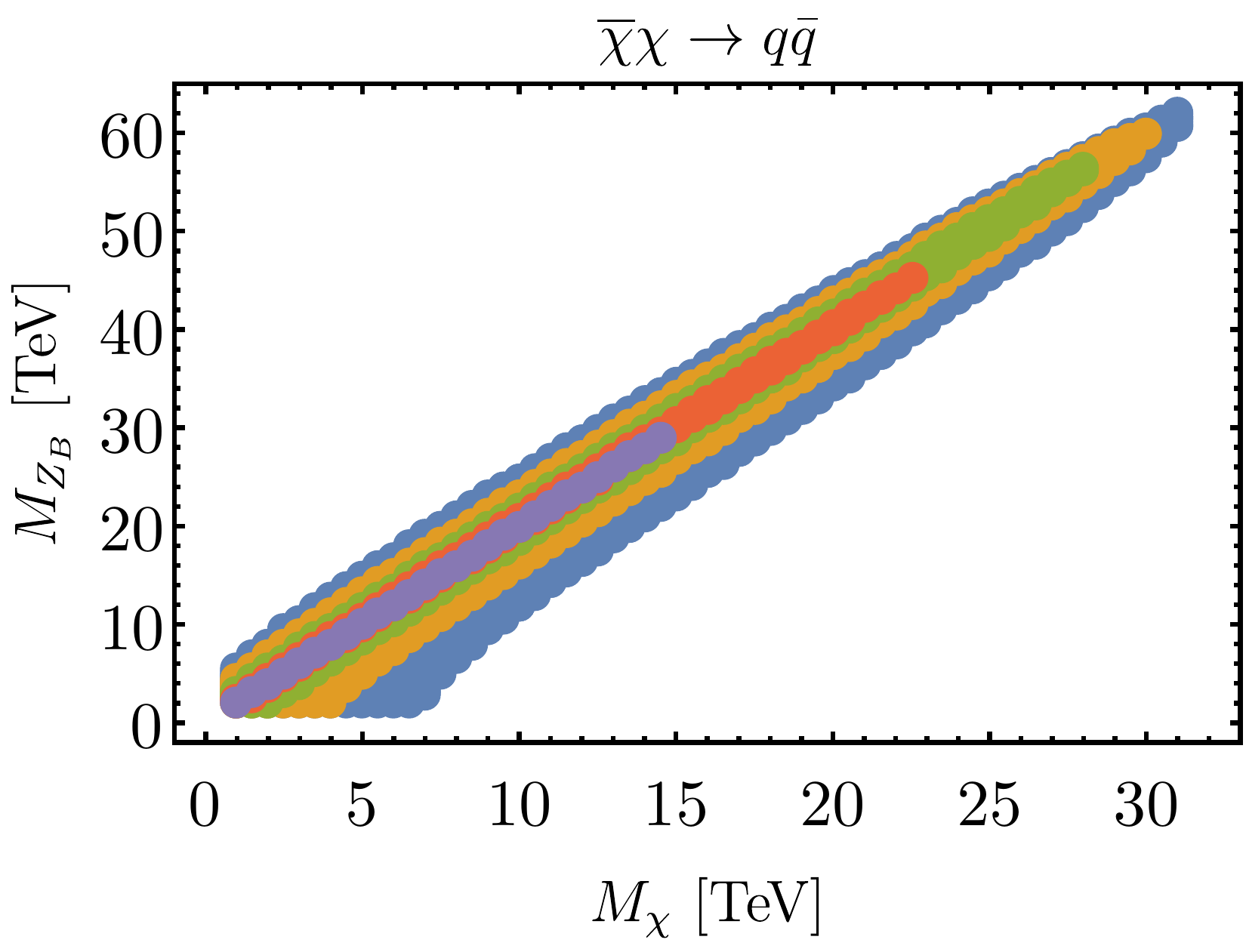} \hfill
\includegraphics[width=0.48\linewidth]{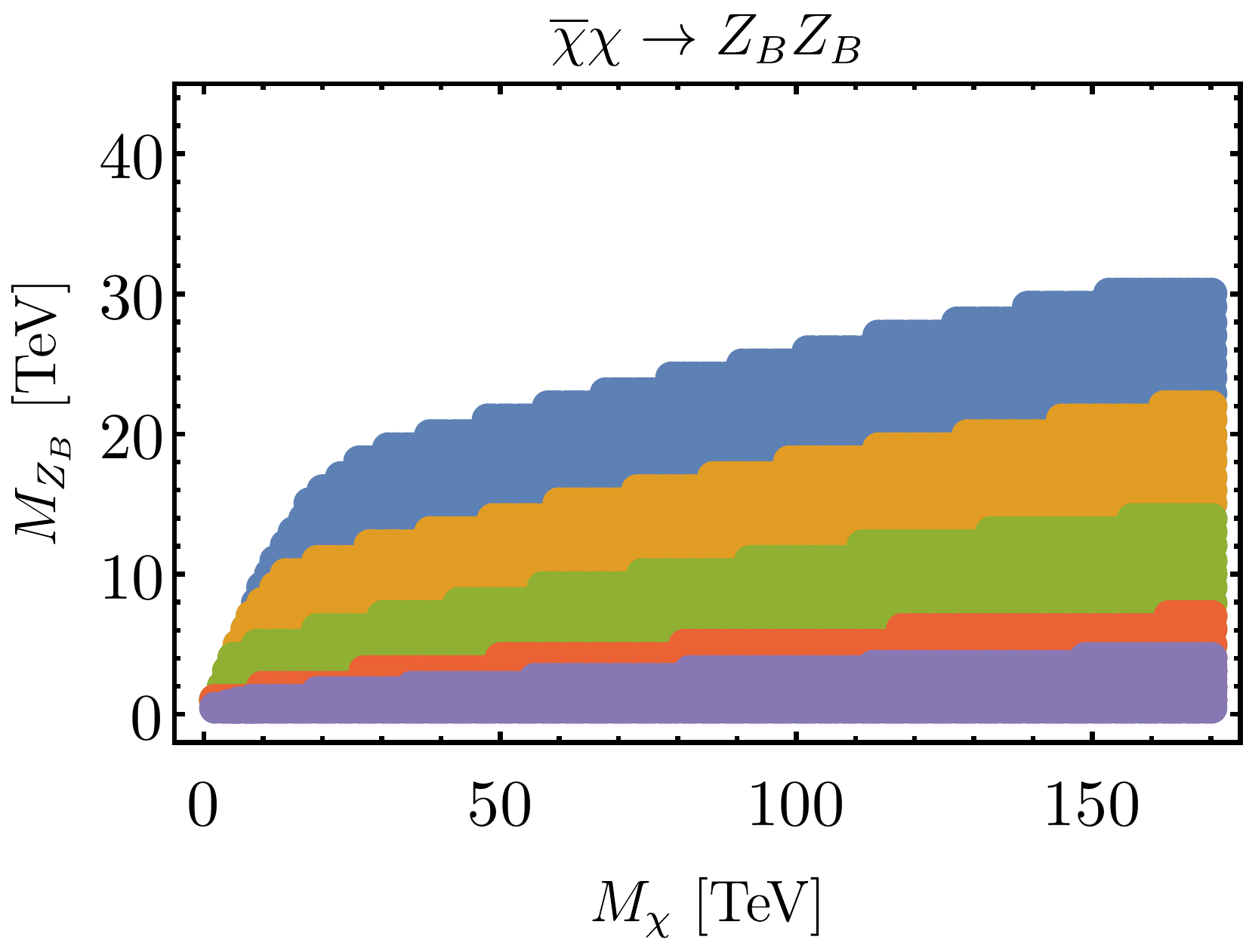} 
\includegraphics[width=0.48\linewidth]{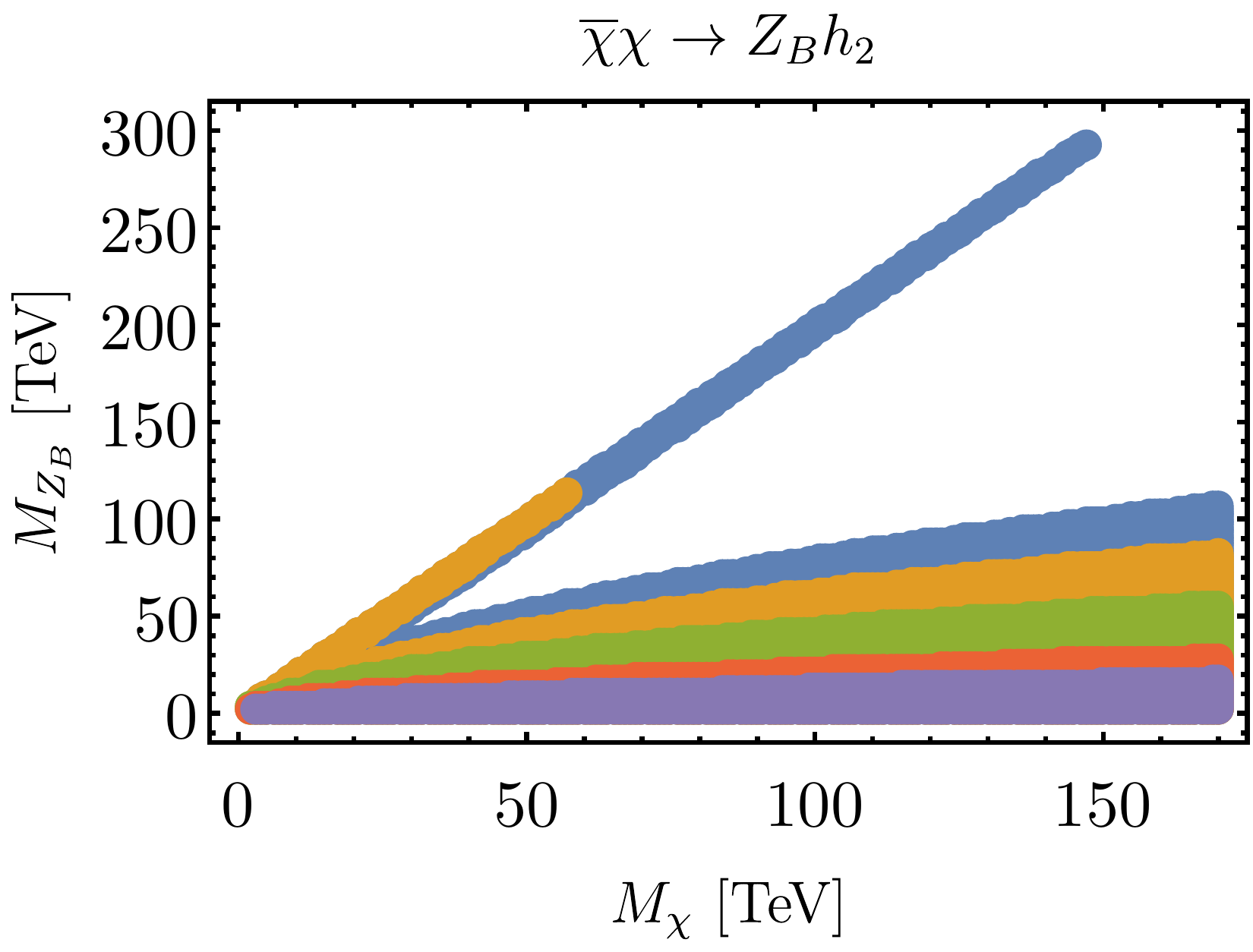} \hfill
\includegraphics[width=0.48\linewidth]{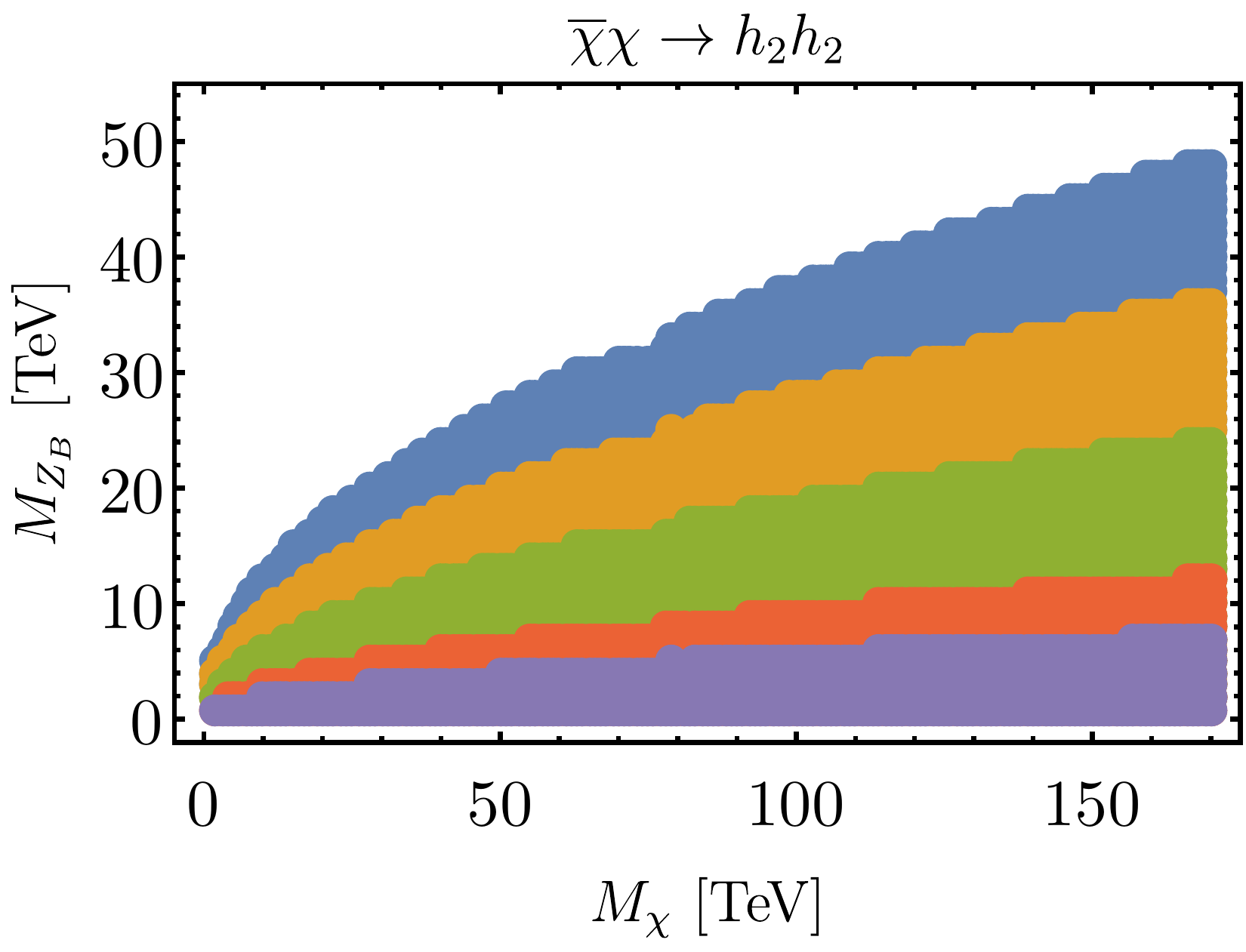} 
\includegraphics[width=0.45\linewidth]{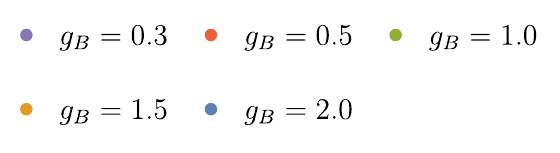} 
\caption{Allowed regions by the cosmological bound on the relic density for each annihilation channel when $\theta_B=0$, and using $M_{h_2}=1$ TeV.}
\label{Annihilation-all-minimal-mixing}
\end{figure}

In Fig.~\ref{Annihilation-all-minimal-mixing}, we show the parameter space in the $M_{Z_B}-M_\chi$ plane allowed by the cosmological constraint $\Omega_{DM} h^2 \leq 0.12$. 
We consider each channel independently to make a detailed discussion. 

\begin{itemize}

\pagebreak

\item Annihilation into two quarks:

In the top-left panel of Fig.~\ref{Annihilation-all-minimal-mixing}, we show the allowed parameter space when we have 
only the annihilation into two quarks: 
\begin{displaymath}
\bar{\chi} \chi \to Z_B^* \to \bar{q} q.
\end{displaymath}
As one would expect, one can sit close to the resonance $M_\chi \sim M_{Z_B}/2$ and achieve a large annihilation cross section, which easily satisfies the bound $\Omega_{DM} h^2 \leq 0.1199 \pm 0.0027$~\cite{Ade:2013zuv}. 
One can see, in Fig.~\ref{Annihilation-all-minimal-mixing}, that the allowed region is in fact around the resonance. 
We note that using the perturbativity bound on the gauge coupling $g_B \leq 2 \sqrt{\pi}$, we find an upper bound on the mass of the leptophobic gauge boson around 65 TeV in this case.

\pagebreak

\item Annihilation into two leptophobic gauge bosons:

In the top-right panel of Fig.~\ref{Annihilation-all-minimal-mixing}, we show the allowed parameter space when one has the annihilation into two leptophobic gauge bosons:
\begin{displaymath}
\bar{\chi} \chi \to Z_B Z_B \  \textrm{(t and u channels), and}  \  \bar{\chi} \chi \to h_2^* \to Z_B Z_B.
\end{displaymath}
In this case, we have the $u$, $t$, and s-channel contributions due to the fact that the new Higgs couples to the dark matter and gauge bosons.
As one can appreciate, one can increase the dark matter mass and find solutions in agreement with the cosmological bound $\Omega_{DM} h^2 \leq 0.12$.
However, as we will discuss later, the perturbativity bound on the Yukawa coupling $\lambda_\chi$ rules out a large fraction of the parameter space for large dark matter mass values and defines an upper bound for the $Z_B$ mass in the context of this annihilation channel.

\item Annihilation into the leptophobic gauge boson $Z_B$ and the new Higgs $h_2$:

In this case, we have three contributions to the annihilation into $Z_B$ and $h_2$: the t- and u-channel contributions, and the s-channel contribution:
\begin{displaymath}
\bar{\chi} \chi \to Z_B h_2 \  \textrm{(t and u channels), and}  \  \bar{\chi} \chi \to Z_B^* \to Z_B h_2.
\end{displaymath}
In the bottom-left panel of Fig.~\ref{Annihilation-all-minimal-mixing}, one can see that there are two main regions in agreement with cosmology. As one can appreciate, in the region where $M_\chi \gg M_{Z_B}$ there is like a plateau for the $Z_B$ mass, while in the second region $2 M_\chi \sim M_{Z_B}$, a portion of parameter space is allowed near the resonance. In both cases one can find an upper bound on the symmetry breaking scale as we will discuss later. 

\item Annihilation into two new Higgses $h_2$:

One has also three type of contributions for the annihilation into two Higgses, we have the u and t channels, and the s-channel mediated by the Higgs boson:
\begin{displaymath}
\bar{\chi} \chi \to h_2 h_2 \  \textrm{(t and u channels) and}  \  \bar{\chi} \chi \to h_2^* \to h_2 h_2.
\end{displaymath}
In the bottom-right panel of Fig.~\ref{Annihilation-all-minimal-mixing}, we show the numerical results to understand the region of the parameter space allowed by cosmology. 
As we will explain later, using the perturbativity bound on the Yukawa coupling $\lambda_\chi < 4 \pi$, we can find an upper bound on the symmetry breaking scale. 
\end{itemize}

Now, combining all the above annihilation channels, we can show the full parameter space allowed by cosmology. 
Furthermore, it is important to use the perturbativity bound on the relevant Yukawa couplings. In this case we can write 
\begin{displaymath}
\lambda_\chi = \frac{3 \sqrt{2} g_B M_\chi}{M_{Z_B}} \leq 4 \pi.
\end{displaymath}
The perturbativity bound on $\lambda_\chi$ is crucial to find the allowed region in this model. In Fig.~\ref{Full-minimal-mixing}, we show the parameter space 
allowed by the relic density constraints and perturbativity including all annihilation channels when $\theta_B=0$. We take $M_{h_2}=1$ TeV in order to be conservative since, as we will discuss later, the upper bound reaches its largest value in this case. On the other hand, we also show in the figure the allowed parameter space by the unitarity bound on the S-matrix. As Fig.~\ref{Full-minimal-mixing} shows, the unitarity bound reduces the upper bound given by the relic density constraint to, approximately, 200 TeV.  Therefore, there is clearly an upper bound on the symmetry breaking scale, and for this channel it is around 200 TeV. We refer the reader to Appendix C for a detailed discussion on the unitarity bounds.
%
\begin{figure}[h]
\centering
\includegraphics[width=0.75\linewidth]{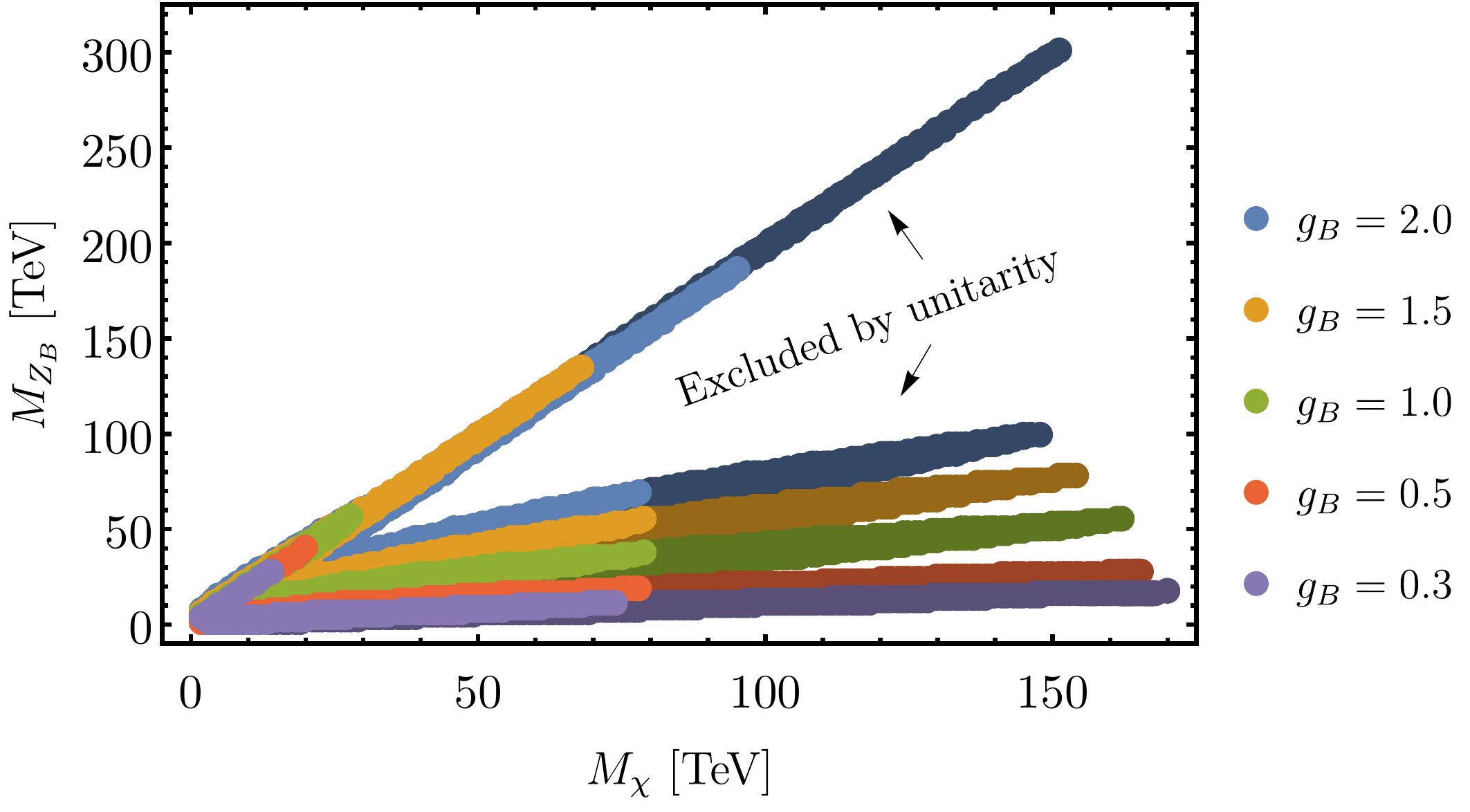}  
\caption{Parameter space allowed by the relic density constraint and perturbativity including all annihilation channels when $\theta_B=0$, using $M_{h_2}=1$ TeV. The dark regions are excluded by unitarity, see Appendix C for details.}
\label{Full-minimal-mixing}
\end{figure}

\item Maximal Mixing Scenario

The mixing angle between the two Higgses can be as large as $\theta_B=0.36$, and, in this case, there are more relevant annihilation channels. The dark matter annihilation channels 
are
$$\bar{\chi} \chi \to \bar{q} q, WW, ZZ, h_1 h_1, Z_B Z_B, Z_B h_2, Z_B h_1, h_2 h_2, h_1 h_2.$$
In order to understand the importance of the different channels, we plot the branching ratios for the different dark matter annihilation 
channels when the mixing angle is $\theta_B=0.36$ in Fig.~\ref{Brs-Maximal}. In this case, we use the following values for the input parameters: $M_{Z_B}=3$ TeV, $M_{h_2}=1$ TeV, $g_B=0.5$, $x_f=24$, and $B=-1$.
We note that, around the resonance (i.e. $M_\chi \sim 1.5$ TeV), the annihilation into two quarks is very important, and as the dark matter mass gets closer to the $Z_B$ mass (i.e. above 2 TeV), the annihilation channel 
$\bar{\chi} \chi \to Z_B h_2$ dominates. Therefore, we can say that this channel is crucial to find the upper bound on the symmetry breaking scale in both scenarios.
\begin{figure}[h]
\includegraphics[width=0.8\linewidth]{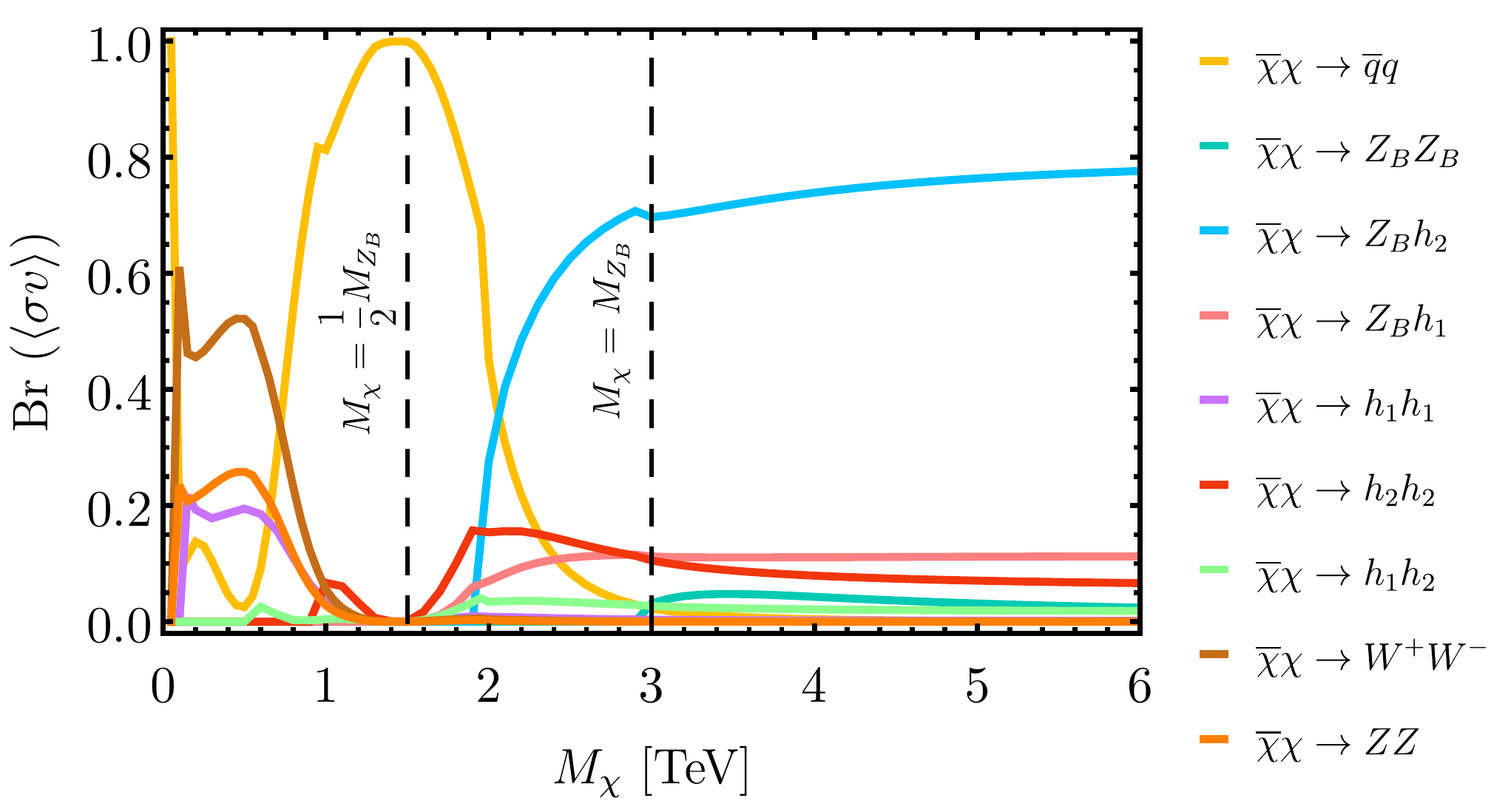} 
\caption{Branching ratios for the different dark matter annihilation channels when the mixing angle is $\theta_B=0.36$. For illustration, we use the following values for the input parameters: 
$M_{Z_B}=3$ TeV, $M_{h_2}=1$ TeV, $g_B=0.5$, $x_f=24$, and $B=-1$.}
\label{Brs-Maximal}
\end{figure}
In Fig.~\ref{Full-maximal-mixing}, we show the parameter space allowed by the relic density constraint and perturbativity including all annihilation channels when $\theta_B=0.36$. 
For illustration, we have taken $M_{h_2}=1$ TeV. As in the case of zero mixing 
angle, the annihilation channel $\bar{\chi} \chi \to Z_B h_2$ defines the upper bound on the symmetry breaking scale, and in this case the maximal allowed value for $M_{Z_B}$ is 
slightly above 200 TeV, very similar to the zero mixing angle scenario, taking also into account the unitarity bound of the S-matrix as mentioned in the previous case.
\begin{figure}[h]
\includegraphics[width=0.75\linewidth]{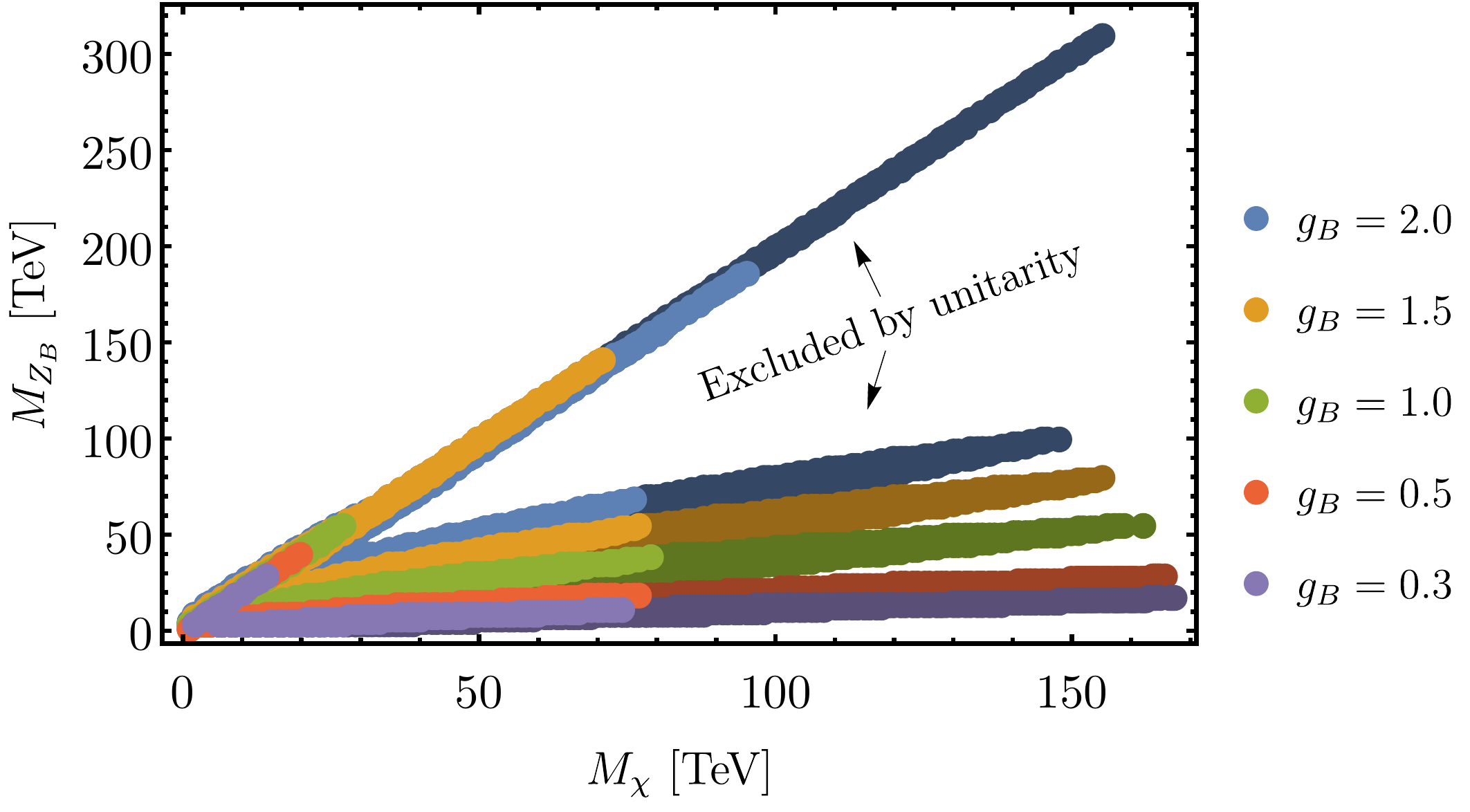} 
\caption{Parameter space allowed by the relic density constraint and perturbativity including all annihilation channels when $\theta_B=0.36$, and using $M_{h_2}=1$ TeV. The dark regions are excluded by unitarity, see Appendix C for details.}
\label{Full-maximal-mixing}
\end{figure}
\end{itemize}
%


\subsection{Direct Detection}
\begin{figure}[t]
\centering
\includegraphics[width=0.7\linewidth]{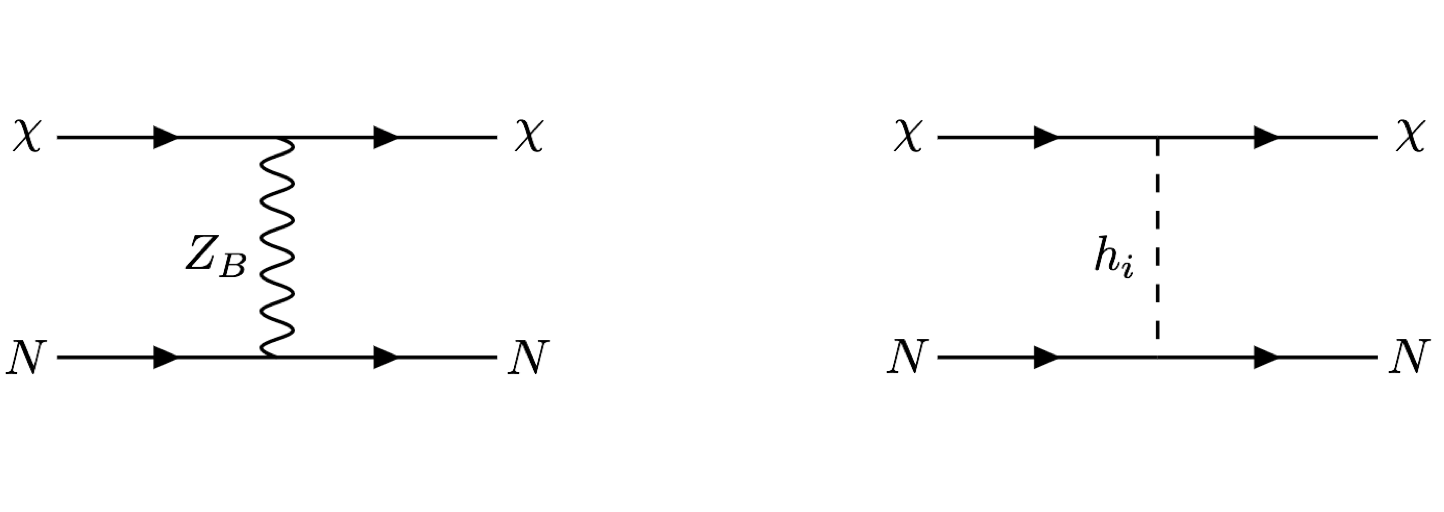} 
\caption{Feynman diagrams relevant for dark matter direct detection.}
\label{DD}
\end{figure}
\begin{figure}[t]
\centering
\includegraphics[width=0.48\linewidth]{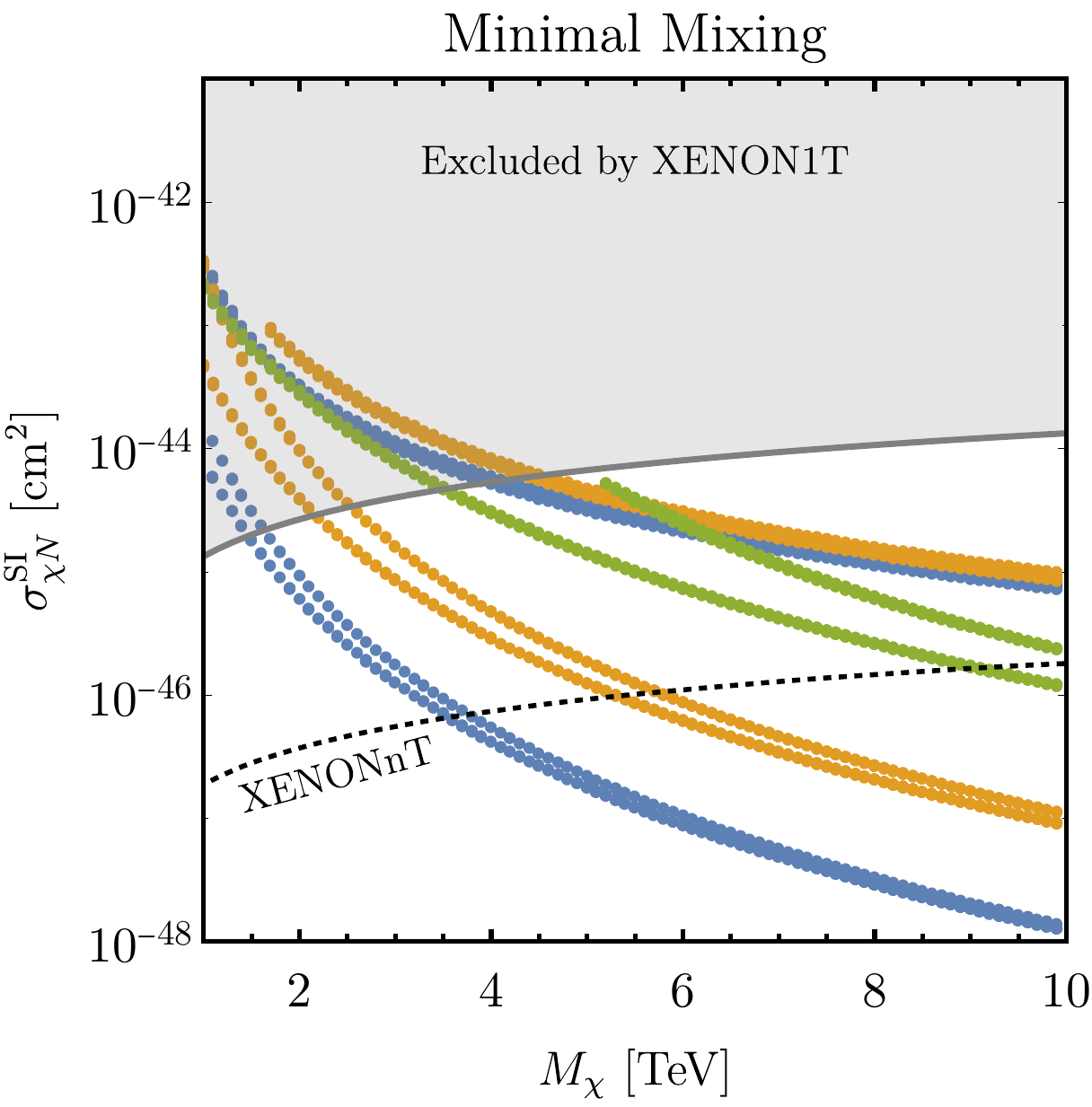} \hfill
\includegraphics[width=0.48\linewidth]{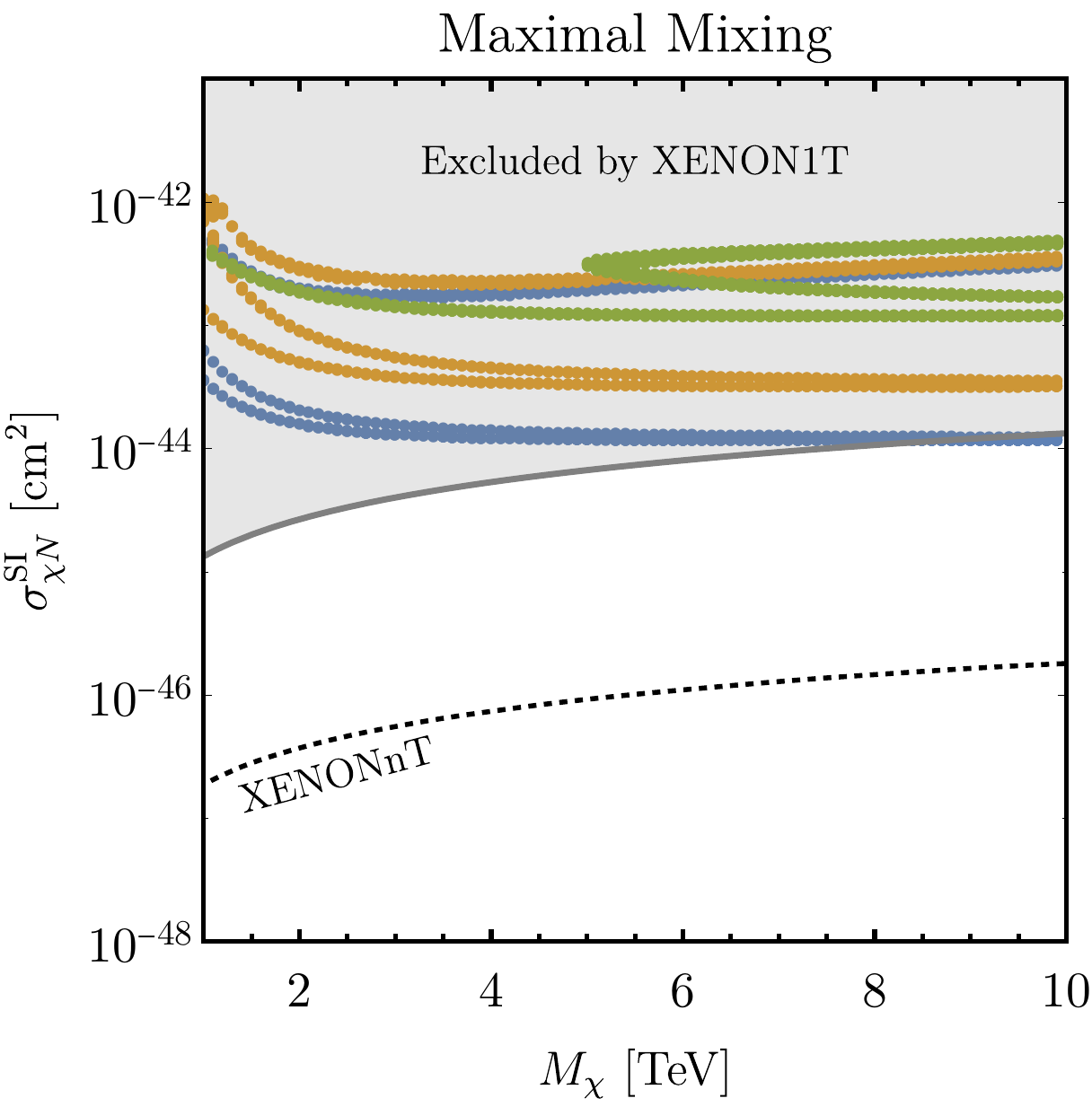} 

\vspace{0.2cm}

\includegraphics[width=0.45\linewidth]{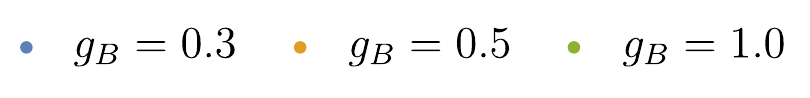} 
\caption{Predictions for the spin-independent dark matter-nucleon scattering cross section in the context of the minimal (left panel) and maximal (right panel) mixing scenarios in agreement with the dark matter relic density constraint. The gray shaded area represents the excluded area by the Xenon1T bounds~\cite{XENON1T}, and the dashed line corresponds to the projected Xenon-nT bounds~\cite{XENONnT}. Notice that there are multiplet curves corresponding to the same color because these regions are allowed by the relic density constraints for a given value of the gauge coupling.}
\label{Direct-detection}
\end{figure}

In the previous study, we have shown that this theory must be realized at the low scale 
in order for the theory to be in agreement with the cosmological bounds on the relic density.
Nevertheless, one has also to take into account an important aspect of any dark matter study: the study of the predictions for the 
direct dark matter experiments. In this theory, the spin-independent elastic 
nucleon-dark matter cross section is given by
\begin{equation}
\sigma_{\chi N}^\text{SI} = \f{g_B^2 M_N^2 M_\chi^2}{4\pi M_{h_1}^4 M_{h_2}^4 M_{Z_B}^4 v_0^2(M_\chi+M_N)^2}\qty[2  B g_B  v_0 M_{h_1}^2 M_{h_2}^2+3 f_N M_\chi M_N M_{Z_B}\sin(2\theta_B)(M_{h_1}^2-M_{h_2}^2)]^2.
\end{equation}
where $M_N$ is the nucleon mass, and $f_N$ is the effective Higgs-nucleon-nucleon coupling. In our numerical results, we use $f_N=0.3$~\cite{Hoferichter:2017olk}. 
See Fig.~\ref{DD} for the relevant Feynman graphs in this context. In Fig.~\ref{Direct-detection}, we show the numerical predictions for the spin-independent 
dark matter-nucleon scattering cross section for the minimal (left panel), and maximal (right panel) mixing scenarios in agreement with the dark 
matter relic density constraint. As one can appreciate from the figure, it is difficult to satisfy the direct detection experimental bounds from the Xenon1T experiment~\cite{XENON1T} in the maximal mixing scenario 
because the contribution to the dark matter-nucleon cross section mediated by the Standard Model Higgs is large. We note that, only when the gauge coupling $g_B$ is smaller than 0.3, and 
when the dark matter mass is smaller than 10 TeV, one can satisfy the experimental bounds. When there is no mixing scenario between the Higgses, see the left panel of Fig.~\ref{Direct-detection}, 
one can easily satisfy the experimental bounds if the dark matter mass is greater than a few TeV. For instance, if $g_B=0.5$ and the dark matter is greater than 2 TeV we 
can satisfy the Xenon1T bounds. Notice that there are multiplet curves corresponding to the same color because these regions are allowed by the relic density 
constraints for a given value of the gauge coupling. We would like to emphasize that $\sigma_{SI}$ does not depend strongly on $M_{h_2}$ since the mixing angle cannot be too large.
%

\subsection{Indirect Detection}
\begin{figure}[h]
\centering
\includegraphics[width=0.65\linewidth]{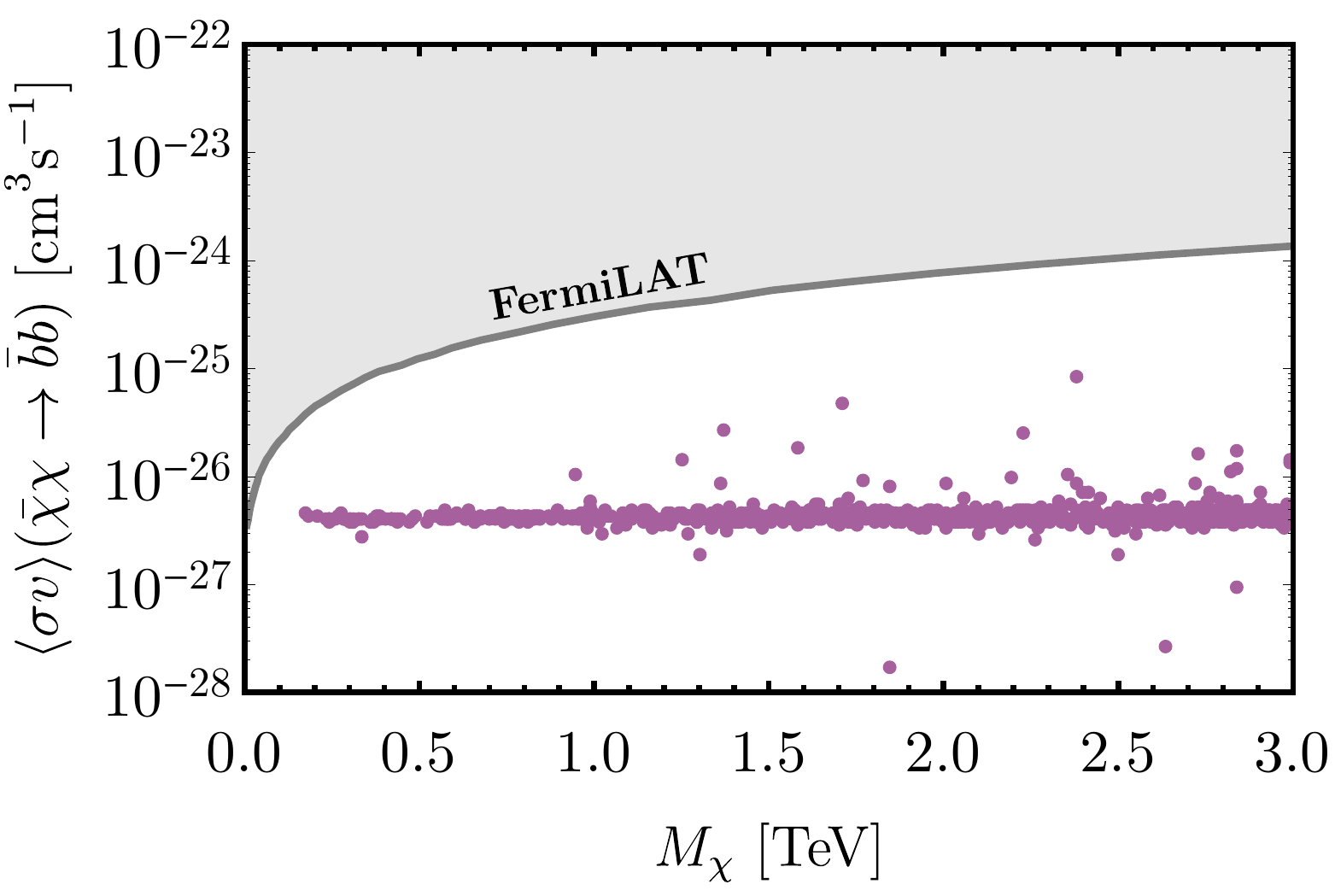} 
\caption{Predictions for the thermal dark matter annihilation into two bottom quarks. In purple we show the points saturating the relic density bound, while the 
gray shaded area shows the parameter space excluded by the Fermi-LAT collaboration~\cite{FermiLAT}.}
\label{ID}
\end{figure}

In this theory, there are several annihilation channels for the leptophobic dark matter candidate, with the annihilation into two bottom quarks when 
$M_\chi \sim M_{Z_B}/2$ being the dominant contribution. Indirect searches by experiments such as Fermi-LAT sets up an upper limit on the thermally averaged cross-section of channels contributing to the photon flux.
In Fig.~\ref{ID}, we show the numerical predictions for $\langle \sigma v \rangle (\bar{\chi}\chi \to \bar{b} b)$ together with the most relevant experimental bound from the Fermi-LAT collaboration~\cite{FermiLAT}. We note that, in the low dark matter mass region, where the experimental bounds become more relevant, the only contribution to the relic density comes from the annihilation into a pair of quarks mediated by the leptophobic gauge boson. Therefore, the predictions shown in Fig.~\ref{ID} depend neither on the choice of the mixing angle $\theta_B$, nor on the choice of the second Higgs mass $M_{h_2}$. As the figure shows, the predictions in this model are compatible with the indirect detection bounds. On the other hand, the dark matter could annihilate through the process $\bar{\chi}\chi \to Z_B h_2$, or we could have gamma-ray lines. Unfortunately, the predictions for the gamma lines are loop suppressed, and it is not possible to distinguish the gamma lines from dark matter annihilation from the continuum spectrum.
%
\subsection{Upper Bound on the Baryon Number Violation Scale}
\begin{figure}[h]
\centering
\includegraphics[width=0.6\linewidth]{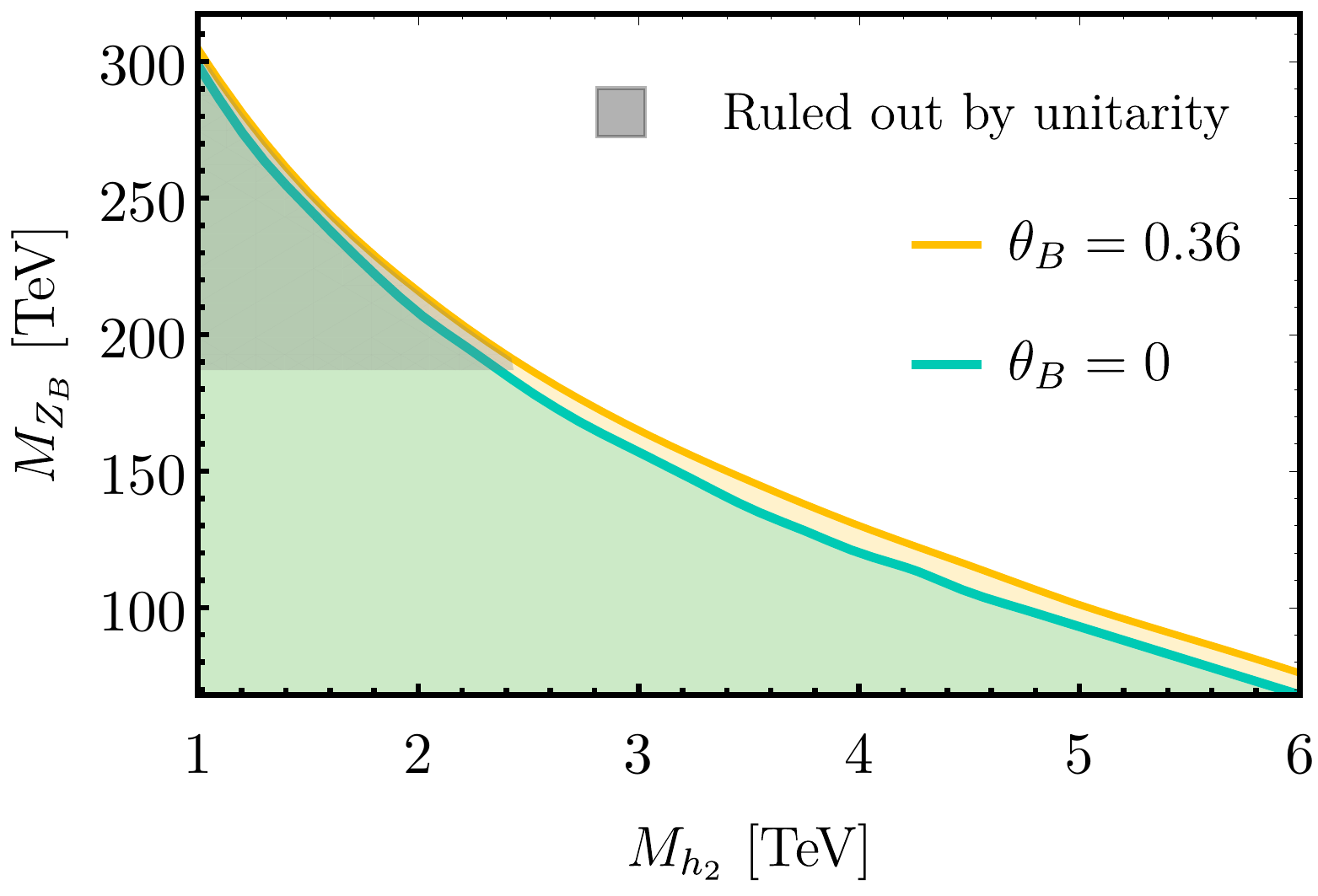} 
\caption{Upper bound on the symmetry breaking scale imposed by meeting the relic density constraint $\Omega_\chi h^2 = 0.12$ 
as a function of the second Higgs mass, for the two extreme scenarios: minimal and maximal mixings.}
\label{UBDM}
\end{figure}
In Figs.~\ref{Full-minimal-mixing} and \ref{Full-maximal-mixing}, we showed the parameter space allowed by the relic density bound for different choices of the parameters of the model. As one can see, the upper bound in this model is defined by the annihilation channel $\bar{\chi} \chi \to Z_B h_2$. For large $M_{h_2}$, the upper bound would be given by the annihilation into a pair of quarks when the resonance is reached, see Fig.~\ref{UBDM} . The annihilation into two gauge bosons, as well as into two Higgses, could be relevant per se, but these channels are bounded by perturbativity of the couplings. We note that, in the context of this model, the dark matter mass and the new gauge boson mass are related as follows
\begin{equation}
M_{Z_B} = \frac{3 \sqrt{2} }{ \lambda_\chi g_B} M_\chi.
\end{equation}
Given a mass for the dark matter candidate, the perturbativity bound on the coupling $\lambda_\chi$ defines the lowest possible mass that the leptophobic gauge boson can have. As seen in Figs. \ref{Full-minimal-mixing} and \ref{Full-maximal-mixing}, this constraint rules out part of the parameter space in the non-resonance region, making the annihilation channels that are relevant near the resonance responsible for the upper bound on the symmetry breaking scale. 
In Fig.~\ref{UBDM}, we show the values for the upper bound imposed by meeting the relic density constraint $\Omega_\chi h^2 = 0.12$ 
as a function of the second Higgs mass in the two scenarios studied above: minimal and maximal mixing scenarios. As we can see in the figure, the upper bound on the symmetry breaking scale 
is around 200 TeV. However, one also has to take into account the bounds coming from the unitarity of the S-matrix which might become relevant for heavy masses. In our case, the unitarity bound is stronger than the bound given by the relic density constraint for $h_2$ masses below 2.5 TeV, and impose an upper bound in that region around 200 TeV. Therefore, we can hope to test this theory at current or future colliders, and there are interesting implications for cosmology; for example, any mechanism for 
baryogenesis should take into account the fact that the local baryon number in this theory is broken below 200 TeV.  

\FloatBarrier

\section{SUMMARY}
In order to investigate the possibility to find the upper bound on the baryon number violation scale, we have investigated the properties of a leptophobic dark matter 
candidate in a simple theory where the local baryon number is broken at the low scale. We have studied all the annihilation channels in great detail and found 
the allowed parameter space in agreement with the cosmological bounds on the cold relic density. Using the cosmological bounds on the relic density, we find that 
the local baryon number symmetry must be broken below the 200 TeV scale. This is a striking result which tells us that this theory could be tested in the near future 
at collider experiments. 

The unitarity constraints are very important in our study. It is well-known, that the unitarity constraints generically imposes an upper bound around 
100 TeV for a thermal produced cold dark matter candidate. However, it is not always the case that the bound coming from unitarity constrains 
the scale of the new theory, i.e. the mass of the new mediator.  In general, it will constrain the ratio between the different mass scales in the theory.
The theory we investigated in this paper is very unique because both, the dark mater and the new gauge boson $Z_{B}$,  acquire their masses from the 
same symmetry breaking scale. Also we should mention that if the theory does not live on the $Z_{B}$ resonance region, the upper bound on the symmetry breaking scale will be around $70$ TeV, 
much smaller than 200 TeV.  The resonance region is always allowed but it is not the most generic scenario. It is important to mention that the region where 
$2 M_{\chi} <  M_{Z_B}$  is not allowed by the perturbativity conditions on the parameters of the scalar potential. Notice that this region of the parameter space 
generically could give a different upper bound but it is not even allowed. Since this model has only a few relevant parameters for our study, 
we were able to perform a general study including all channels and constraints to find the upper bound on the symmetry breaking scale.
All these interesting features tell us that this theory is a good predictive theory for dark matter.

The upper bound on the symmetry breaking scale also has profound implications for cosmology; in particular to baryogenesis, since the scale 
for baryon number violation must be low. We would like to emphasize that this theory does not have the main problem of most of the extensions of the Standard 
Model, where the new physical scale can be very large and one cannot be sure about the possibility to test these theories.

One of the main implications of having a low scale for the spontaneous breaking of local baryon number is that one needs to take into account the fact 
that the local baryon number can be broken at the very low scale. The simplest scenario for baryogenesis in this case is to have leptogenesis at the high scale 
and impose the conditions on the chemical potentials due to the conservation of baryon number. In this case, the lepton asymmetry generated by leptogenesis 
is converted to a baryon asymmetry by the sphalerons, but the conversion factor is smaller than the conversion factor in the Standard Model; see 
Refs.\cite{Perez:2013tea,Duerr:2014wra,Perez:2015rza} for the study of baryogenesis in these theories. With the need to break the local baryon number at the low scale as motivation, we could think in the future about the collider signatures, the study of topological effects, and the study of the phase transitions related to the spontaneous breaking of baryon number in nature.  

\section*{Acknowledgments:}
{\small{P. F. P. thanks Mark B. Wise for discussions and the Walter Burke Institute for Theoretical Physics at Caltech for hospitality and support.
This work has been partially supported by the U.S. Department of Energy, Office of Science, Office of High Energy Physics, under Award Number DE-SC0011632. 
This work made use of the High Performance Computing Resource in the Core Facility for Advanced Research Computing at Case Western Reserve University. The work of C.M. has been supported in part by Grants No. FPA2014-53631-C2-1-P, FPA2017-84445-P and SEV-2014-0398 (AEI/ERDF, EU), and ``La Caixa-Severo Ochoa" scholarship.}}
\appendix
%
\section{Feynman Rules}
\begin{eqnarray}
\Gamma^\mu_{X_i^0 X_j^0 Z} &:&  -\frac{i g_2}{2\cos{\theta_w}}(U_{3i}U_{j3} + U_{4i}U_{j4})\gamma^\mu P_L, \\ 
\Gamma^\mu_{X_i^0 X_j^0 Z_B} &:& -i g_B [B_1(U_{3i}U_{j3} + U_{2i}U_{j2}) + B_2(U_{1i}U_{j1} + U_{4i}U_{j4})]\gamma^\mu P_L, \\
\Gamma_{X_i^0 X_j^0 h_1} &:& -i \sqrt{2} C \left[ \sin \theta_B (\lambda_\chi U_{i2} U_{j1} + \lambda_\Psi U_{i4} U_{j3})  + \cos \theta_B (y_2 U_{i2} U_{j3} + y_4 U_{i4} U_{j1})  \right], \\
\Gamma_{X_i^0 X_j^0 h_2} &:&  -i \sqrt{2} C \left[ \cos \theta_B (\lambda_\chi U_{i2} U_{j1} + \lambda_\Psi U_{i4} U_{j3})  - \sin \theta_B (y_2 U_{i2} U_{j3} + y_4 U_{i4} U_{j1})  \right],  \\
\Gamma_{h_1 Z_B Z_B}^{\mu \nu}  &:& 2i g^{\mu \nu}  \frac{M_{Z_B}^2}{v_B} \sin{\theta_B}, \\
\Gamma_{h_2 Z_B Z_B}^{\mu \nu}  &:&  2i g^{\mu \nu}  \frac{M_{Z_B}^2}{v_B} \cos{\theta_B}, \\
\Gamma_{h_1 Z Z}^{\mu \nu}  &:& 2i g^{\mu \nu}  \frac{M_{Z}^2}{v_0} \cos{\theta_B}, \\
\Gamma_{h_2 Z Z}^{\mu \nu}  &:& -2i g^{\mu \nu}  \frac{M_{Z}^2}{v_0} \sin{\theta_B},  \\
\Gamma_{h_1 W W}^{\mu \nu}  &:& 2i g^{\mu \nu}  \frac{M_{W}^2}{v_0} \cos{\theta_B}, \\
\Gamma_{h_2 W W}^{\mu \nu}  &:& -2i g^{\mu \nu}  \frac{M_{W}^2}{v_0} \sin{\theta_B},  \\
\Gamma_{h_1 h_1 h_1} &:& 3i [ 2  \lambda_H v_0 \cos^3 \theta_B + 2 \lambda_B v_B  \sin^3 \theta_B+\lambda_{HB} (v_0\cos\theta_B\sin^2\theta_B +v_B\cos^2\theta_B\sin\theta_B)], \hspace{1.5cm} \\
\Gamma_{h_1 h_1 h_2} &:& i [ - 6  \lambda_H v_0 \cos^2 \theta_B \sin \theta_B+ 6 \lambda_B v_B \cos \theta_B \sin^2 \theta_B \nonumber+ \\
&& \lambda_{HB} ( v_B \cos^3 \theta_B + 2 v_0 \cos^2 \theta_B \sin \theta_B - 2 v_B \cos \theta_B \sin^2 \theta_B - v_0 \sin^3 \theta_B ) ],\\
\Gamma_{h_2 h_2 h_1} &:& i [ 6 \lambda_H v_0 \cos \theta_B \sin^2 \theta_B+ 6 \lambda_B v_B \cos^2 \theta_B \sin \theta_B \nonumber+ \\
&& \lambda_{HB} ( v_B \sin^3 \theta_B - 2 v_0 \cos \theta_B \sin^2 \theta_B - 2 v_B \cos^2 \theta_B \sin \theta_B + v_0 \cos^3 \theta_B ) ],\\
\Gamma_{h_2 h_2 h_2} &:& 3i [ -2  \lambda_H v_0 \sin^3 \theta_B + 2 \lambda_B v_B  \cos^3 \theta_B+\lambda_{HB} (v_B\cos\theta_B\sin^2\theta_B -v_0\cos^2\theta_B\sin\theta_B)], \\
\Gamma_{\overline{q} q Z_B}^\mu &:& - i \frac{1}{3} g_B \gamma^\mu.
\end{eqnarray}

\newpage
\section{$SU(2)_L$ Multiplet as Dark Matter Candidate}
\begin{figure}[h]
\centering
\includegraphics[width=0.7\linewidth]{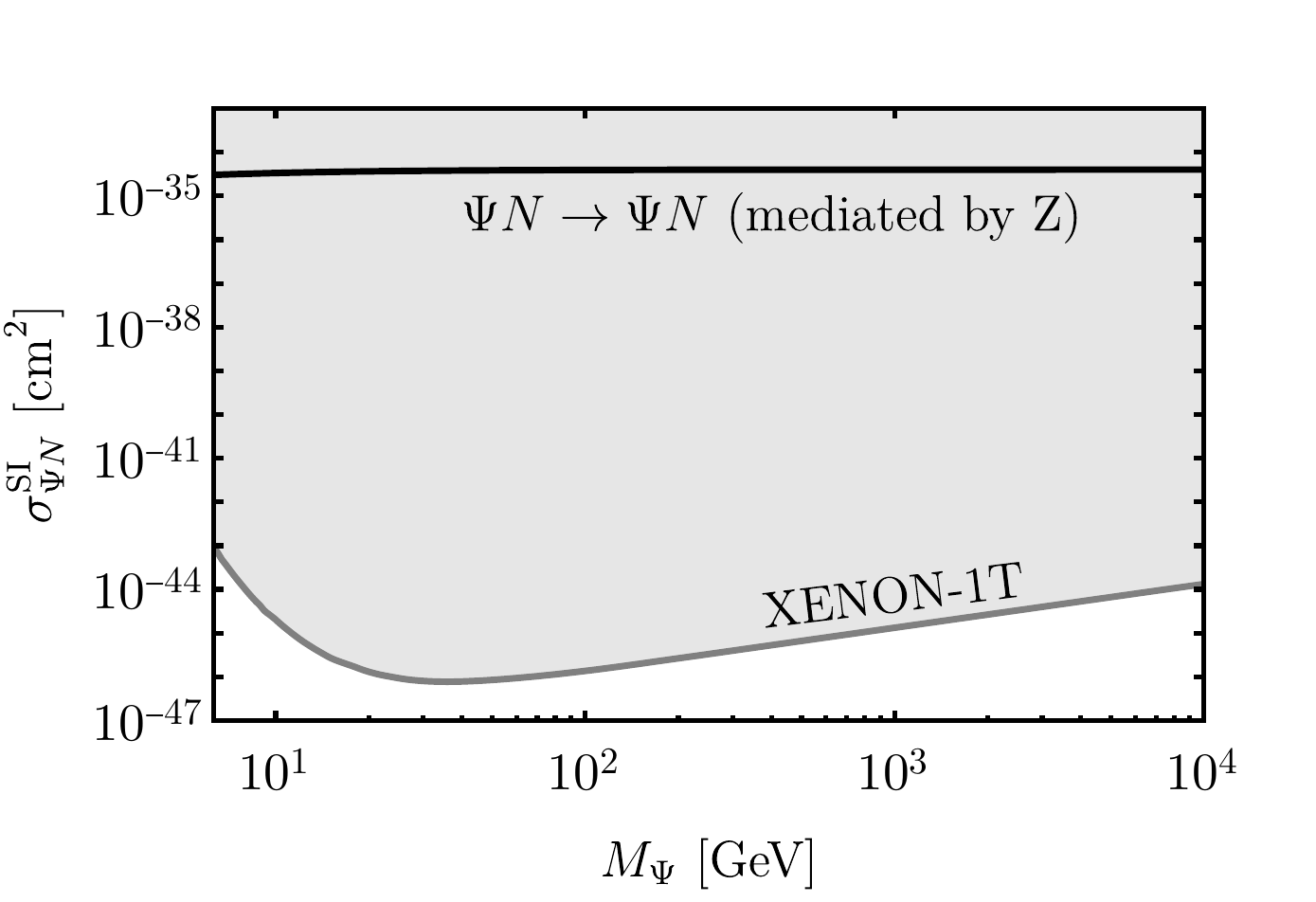}  
\caption{Predictions for the direct detection spin-independent cross section $\sigma^{\text{SI}}_{\Psi N}$. The XENON-1T bounds rule out the gray shaded region.}
\label{DDPsiPlot}
\end{figure}

In our discussion, we have assumed that the lightest new neutral fields corresponds to the field $\chi$, 
and one could wonder whether the neutral $\Psi=\Psi_L + \Psi_R$ could be as well a 
viable dark matter candidate. However, as we show in this section, this possibility is ruled 
out by the direct detection bounds because $\Psi$ has an unsuppressed coupling to the $Z$ gauge boson.
The neutral field $\Psi$ can interact with nuclei through processes mediated by both $Z$ and $Z_B$ bosons. 
We will focus on the contribution mediated by $Z$ since it totally dominates the scattering. 
The relevant Feynman rules for this process are given by: 
\begin{eqnarray}
\Gamma^\mu_{\bar{\Psi}\Psi Z} &:& -i\frac{g_2}{2\cos \theta_W}\gamma^\mu, \\
\Gamma^\mu_{\bar{q}Zq} &:& -i\frac{g_2}{2\cos \theta_W}\gamma^\mu(c_V^q+c_A^q\gamma_5),
\end{eqnarray}
where
\begin{align*}
&c_V^u= \frac{1}{2}-\frac{4}{3}\sin^2 \theta_W, \,\,\,\,\,\,\,\,\,\,  c_A^u=\frac{1}{2}, \\
&c_V^d=-\frac{1}{2}+\frac{2}{3}\sin^2\theta_W, \,\,\,\,\,  c_A^d=-\frac{1}{2}.
\end{align*}
The amplitude for the quark-dark matter spin-independent elastic scattering ($q^2 \to 0$) is given by 
\begin{equation}
{\cal M}= \frac{g_2^2}{4\cos \theta_W^2}\frac{c_V^q}{M_Z^2} (\bar{\Psi}\gamma^\mu \Psi)(\bar{q}\gamma_\mu q).
\end{equation}
Now, we can write the amplitude for the nucleon-dark matter spin-independent elastic scattering as
\begin{eqnarray}
{\cal M}&=& \frac{g_2^2}{4\cos \theta_W^2}\frac{c_V^q}{M_Z^2} (\bar{\Psi}\gamma^\mu \Psi) \langle N | (\bar{q}\gamma_\mu q) | N \rangle,\\
&=& \frac{g_2^2}{4\cos \theta_W^2}\frac{c_V^q}{M_Z^2} (\bar{\Psi}\gamma^\mu \Psi) \left( Z \langle p | (\bar{q}\gamma_\mu q) | p \rangle +(A-Z)  \langle n | (\bar{q}\gamma_\mu q) | n \rangle \right),
\end{eqnarray}
where $Z$ and $A$ are the atomic and mass numbers, respectively. Here,
\begin{equation}
\langle N | \bar{q} \gamma^\mu q | N \rangle = \bar{u}_N \left( \gamma^\mu F_1(q^2)+i\frac{\sigma^{\mu \nu}q_\nu}{2M_N}F_2(q^2) \right) u_N,
\end{equation}
where $N=n,p$, and $F_1(q^2)$ and $F_2(q^2)$ are form factors that only depend on the transferred momentum $q^2$. In the limit of low $q^2$, the only contribution is vectorial and, since $F_1(0)=1$, $\Gamma^\mu \sim \gamma^\mu$. Therefore, at zero momentum transfer, only valance quarks in the nucleon contribute to the vector currents, and the nuclear amplitude reads as
\begin{equation}
{\cal M}= \frac{g_2^2}{4\cos \theta_W^2}\frac{c_V^N}{M_Z^2} (\bar{\Psi}\gamma^\mu \Psi)(\bar{u}_N \gamma_\mu u_N),
\end{equation}
where
\begin{equation}
c_V^N = Z (2 c_V^u + c_V^d) + (A-Z) (c_V^u + 2c_V^d),
\end{equation}
Taking the non relativistic limit of the dark matter candidate spinor,
\begin{equation}
\bar{\Psi} \gamma^\mu \Psi \to 2 M_\Psi \delta^{\mu 0},
\end{equation}
the squared amplitude reads as
\begin{equation}
{\abs{\cal M}}^2 = \frac{g_2^4}{ \cos^4 \theta_W} \frac{ |c_V^N|^2 }{M_Z^4} M_N^2 M_\Psi^2.
\end{equation}
The above amplitude defines the spin-independent cross-section of the process, which is given by 
\begin{equation}
\sigma^{\text{SI}}_{\Psi N}=\frac{g_2^4}{ 16 \pi \cos^4 \theta_W} \frac{ |c_V^N|^2 }{M_Z^4} \mu_{\Psi N}^2,
\end{equation}
where
\begin{equation}
 \mu_{\Psi N} = \frac{M_N  M_\Psi}{M_N+M_\Psi}.
 \end{equation}
In Fig.~\ref{DDPsiPlot}, we show the prediction for the spin-independent cross section of the dark matter scattering with nuclei. In this case, liquid Xenon with numbers $Z = 54$, and $A = 131$ is used. As it can be seen in the figure, the experimental bounds from XENON-1T are many orders of magnitude stronger than the theoretical prediction, and therefore, the possibility of having $\Psi$ as a dark matter candidate is ruled out. 

We note that our main goal in this appendix is to show that the $SU(2)_L$ multiplet $\Psi$ has a huge cross-section, and this is why we focus on the case where the dark matter candidate 
is defined by the properties of the $\chi$ field. Of course, we could consider mixing between between these two fields, but clearly the mixing has to be very small to satisfy 
the experimental bounds.

\newpage

\section{Unitarity Constraints on Dark Matter Mass}
Here we revisit the bound on the dark matter mass pointed out in Ref.~\cite{KamionkowskiGriest:1990}. Starting with the unitarity of the S-matrix, $S^\dag S = 1$, and $S = 1+i T$, one finds
\be \label{C1}
i(T^\dag - T) = T^\dag T.
\ee 
Consider a general scattering process from the initial state $\ket{\alpha}$ to the final state $\ket{\beta}$. Using the definition of the $T$ matrix elements $\mel{\beta}{T}{\alpha} = (2\pi)^4\delta^{(4)}(p_\beta-p_\alpha)\mathcal M(\alpha\to \beta)$, we can write the matrix element of the left-hand side of Eq. \eqref{C1} as
\be
i\mel{\beta}{T^\dag-T}{\alpha} = i(2\pi)^4\delta^{(4)}(p_f-p_i)\qty[\mathcal M^*(\beta\to \alpha)-\mathcal M(\alpha\to \beta)].
\ee
Inserting a complete set of intermediate states, 
\be
1 = \sum_\gamma \int \dd\Pi_\gamma \ket{\gamma}\bra{\gamma}= \sum_\gamma \int \prod_i \f{\dd^3p_{\gamma_i}}{(2\pi)^3 2E_{\gamma_i}}\ket{\qty{\gamma_i}}\bra{\qty{\gamma_i}},
\ee
to the matrix element of the right-hand side:
\be
\begin{split}
	\mel{\beta}{T^\dag T}{\alpha} &= \sum_\gamma \int\dd\Pi_\gamma \mel{\beta}{T^\dag}{\gamma}\mel{\gamma}{T}{\alpha}\\
	&=(2\pi)^8\delta^{(4)}(p_\alpha - p_\gamma)\delta^{(4)}(p_\gamma-p_\beta)\qty[\sum_\gamma \int\dd\Pi_\gamma\mathcal M^*(\beta\to \gamma)\mathcal M(\alpha\to\gamma)],
\end{split}
\ee
one obtains the generalized optical theorem:
\be\label{C5}
\mathcal M(\alpha\to\beta)-\mathcal M^*(\beta\to\alpha) = i(2\pi)^4\sum_\gamma\int\dd\Pi_\gamma\delta^{(4)}(p_\alpha-p_\gamma)\mathcal M(\alpha\to\gamma)\mathcal M^*(\beta\to \gamma).
\ee
If $\ket{\alpha}$ is a two-particle state, which is the case for all the dark matter annihilations we consider,  the cross section in the center-of-mass (CM) frame is given by,
\be
\sigma (\alpha\to \gamma) = \f{1}{4 E_\text{CM}\abs{\vec p_i}}\int\dd\Pi_\gamma (2\pi)^4\delta^{(4)}(p_\alpha-p_\gamma)\abs{\mathcal M(\alpha\to \gamma)}^2.
\ee
Then it is easy to see that the generalized optical theorem \eqref{C5} can be cast in the form:
\be \label{C7}
\text{Im}\mathcal M(\alpha\to\alpha) = 2E_\text{CM}\abs{\vec p_i}\sum_\beta\sigma(\alpha\to\beta),
\ee
which implies that the imaginary part of the forward scattering amplitude is proportional to the total scattering cross section. Furthermore, if the final state $\ket\beta$ is also a two-particle state, we can rewrite the 2-body phase space of the final state in the CM frame as
\be
\dd\Phi_\beta = (2\pi)^4\delta^{(4)}(p_\alpha - p_\beta)\f{\dd^3p_{\beta_1}}{(2\pi)^3 2E_{\beta_1}}\f{\dd^3p_{\beta_2}}{(2\pi)^3 2E_{\beta_2}} = \f{1}{16\pi^2}\f{\abs{\vec p_f}}{E_\text{CM}}\dd\Omega_\text{CM},
\ee
where $\vec p_{\beta_1} = -\vec p_{\beta_2} \equiv \vec p_f$ in the CM frame. Therefore, the cross section becomes
\be
\sigma(\alpha\to\beta) = \f{1}{4E_\text{CM}\abs{\vec p_i}}\int\f{\abs{\vec p_f}\dd\Omega_\text{CM}}{16\pi^2 E_\text{CM}} \abs{\mathcal M(\alpha\to\beta)}^2= \int\dd\Omega_\text{CM}\abs{f(\alpha\to\beta)}^2,
\ee
where we define a dimensionfull scattering amplitude,
\be
f(\alpha\to\beta) \equiv \sqrt{\f{\abs{\vec p_i}}{\abs{\vec p_f}}}\f{\mathcal M(\alpha\to\beta)}{8\pi E_\text{CM}}.
\ee
Finally, we obtain the optical theorem in a form that is more familiar to us from quantum mechanics,
\be\label{C11}
\text{Im} f(\alpha\to\alpha) = \f{\abs{\vec p_i}}{4\pi}\sum_\beta \sigma(\alpha\to\beta).
\ee
Now, using the Legendre polynomials we can write the amplitude as
\be\label{C13}
f(\alpha \to \beta) =\sum_J (2J+1)P_J(\cos{\theta})a_J (\alpha \to \beta).
\ee
Writing the total annihilation cross section for the dark matter particles ($\alpha$) into any two-body state ($\beta$) in terms of the partial wave expansion in Eq. \eqref{C13} and using the orthogonality relation of the Legendre polynomials, we obtain
\be \label{C14}
\begin{split}
	\sigma(\alpha \to \beta) &= \int \dd\Omega \abs{f(\alpha \to \beta)}^2 \\
	&= \int \dd\Omega \sum_{J, J'} (2J+1)(2J'+1)P_J(\cos{\theta})P_{J'}(\cos{\theta}) a_J(\alpha \to \beta)a^*_{J'}(\alpha \to \beta) \\
	&=\sum_J 4\pi (2J+1) \abs{a_J(\alpha \to \beta)}^2\\
	&\equiv\sum_J \sigma_J,
\end{split}
\ee
where $\sigma_J\equiv 4\pi(2J+1)\abs{a_J(\alpha \to \beta)}^2$. Using Eqs. \eqref{C13} and \eqref{C14} in the optical theorem \eqref{C11}, we establish the following relation between the partial wave amplitudes,
\begin{eqnarray}\label{C15}
	\f{\text{Im}\,a_J(\alpha\to\alpha)}{\abs{\vec p_i}} = \sum_\beta \abs{a_J(\alpha\to\beta)}^2 = \abs{a_J(\alpha\to\alpha)}^2+\sum_{\beta\neq\alpha}\abs{a_J(\alpha\to\beta)}^2,
\end{eqnarray}
which implies the inequality
\be
\abs{a_J(\alpha \to \alpha)}^2 \leq \frac{\text{Im}\,a_J(\alpha \to \alpha)}{\abs{\vec p_i}},\quad \forall J.
\ee
Hence, 
\begin{eqnarray}
	\qty(\text{Re} \, a_J(\alpha \to \alpha))^2 + \qty( \text{Im} \, a_J(\alpha \to \alpha) - \frac{1}{2\abs{\vec p_i}})^2 \leq \frac{1}{4\abs{\vec p_i}^2},
\end{eqnarray}
and this inequatlity bounds the imaginary part of the elastic scattering partial wave amplitude as
\begin{eqnarray}
	\text{Im}\,a_J(\alpha \to \alpha) \leq \frac{1}{\abs{\vec p_i}}.
\end{eqnarray}
Applying this inequality to Eq. \eqref{C15} gives
\begin{eqnarray}
	\sum_\beta \abs{a_J (\alpha \to \beta)}^2 \leq \frac{1}{\abs{\vec p_i}^2},
\end{eqnarray}
which leads to a constraint on the $J$th partial wave cross section defined in Eq. \eqref{C14} as follows
\begin{eqnarray}
	\sigma_J \leq \frac{4 \pi  (2J + 1)}{\abs{\vec p_i}^2}.
\end{eqnarray} 
Since the dark matter candidate is nonrelativistic, we may approximate the dark matter momentum
\begin{eqnarray}
	\abs{\vec p_i} \simeq \frac{M_\chi v_\text{rel}}{2}
\end{eqnarray}
and hence, the bound on each partial wave spin-averaged cross section is 
\begin{eqnarray} \label{C22}
	\bar{\sigma}_J \lesssim \frac{4\pi}{M_\chi^2 v_\text{rel}^2}(2J+1).
\end{eqnarray}

Since the angular dependence of the cross section arises through the Mandelstam variable $t$, approximating $t$ to lowest order in $v_\text{rel}$ for a dark matter annihilation process $\overline{\chi} \chi \to a b$ gives
\be \label{C23}
\begin{split}
	t &= M_\chi^2 + M_a^2 - 2 E_\chi E_a + 2\abs{\vec p_a} \cos{\theta}\f{M_\chi v_\text{rel}}{2}+\mathcal O\qty(v_\text{rel}^2)\\
	&\simeq M_\chi^2 + M_a^2 - 2 E_\chi E_a.
\end{split}
\ee

So, this approximation results in the cross section with no angular dependence, which corresponds to the $J=0$ partial wave and the unitarity constraint is given by
\begin{eqnarray}
	\bar{\sigma}_0 \leq \frac{4\pi}{M_\chi^2 v_{rel}^2}.
\end{eqnarray}

To implement this constraint, we calculate the total annihilation cross section of the dark matter candidate with the Mandelstam variable $t$ approximated by Eq. \eqref{C23}. Notice that the cross section still depends on the Mandelstam variable $s$, and to avoid the pole in the cross section which arises when $s = 4 M_\chi^2$, we approximate it as $s\simeq 4 M_\chi^2 + M_\chi^2v_\text{rel}^2$. Setting $v_\text{rel} \simeq (6/x_f)^{1/2}$, where $x_f = M_\chi/T_f$, with $T_f$ the freeze-out temperature of the dark matter, we can exclude part of the region in the $M_{Z_B}-M_\chi$ plane that is violating the unitarity constraint, as shown in Figs.~\ref{Full-minimal-mixing} and~\ref{Full-maximal-mixing}.




\begin{thebibliography}{99}

\bibitem{Bertone:2004pz}
  G.~Bertone, D.~Hooper and J.~Silk,
  ``Particle dark matter: Evidence, candidates and constraints,''
  Phys.\ Rept.\  {\bf 405} (2005) 279.
  [hep-ph/0404175].


\bibitem{Pais:1973mi}
  A.~Pais,
  ``Remark on baryon conservation,''
  Phys.\ Rev.\ D {\bf 8} (1973) 1844.

\bibitem{Carone:1995pu} 
  C.~D.~Carone and H.~Murayama,
  ``Realistic models with a light U(1) gauge boson coupled to baryon number,''
  Phys.\ Rev.\ D {\bf 52}, 484 (1995).
  [hep-ph/9501220].
  
  
\bibitem{FileviezPerez:2010gw}
  P.~Fileviez P\'erez and M.~B.~Wise,
  ``Baryon and lepton number as local gauge symmetries,''
  Phys.\ Rev.\ D {\bf 82} (2010) 011901
   [Erratum-ibid.\ D {\bf 82} (2010) 079901].
  [arXiv:1002.1754 [hep-ph]].
  
  
\bibitem{FileviezPerez:2011pt}
  P.~Fileviez P\'erez and M.~B.~Wise,
  ``Breaking Local Baryon and Lepton Number at the TeV Scale,''
  JHEP {\bf 1108} (2011) 068
  [arXiv:1106.0343 [hep-ph]].
  
  
\bibitem{Duerr:2013dza}
  M.~Duerr, P.~Fileviez Perez and M.~B.~Wise,
  ``Gauge Theory for Baryon and Lepton Numbers with Leptoquarks,''
  Phys.\ Rev.\ Lett.\  {\bf 110} (2013) 231801
  [arXiv:1304.0576 [hep-ph]].
  
  
\bibitem{Perez:2014qfa}
  P.~Fileviez Perez, S.~Ohmer and H.~H.~Patel,
  ``Minimal Theory for Lepto-Baryons,''
  Phys.\ Lett.\ B {\bf 735} (2014) 283
  [arXiv:1403.8029 [hep-ph]].


 
\bibitem{Perez:2015rza}
  P.~Fileviez Perez,
  ``New Paradigm for Baryon and Lepton Number Violation,''
  Phys.\ Rept.\  {\bf 597} (2015) 1
  [arXiv:1501.01886 [hep-ph]].
  
 
\bibitem{Duerr:2013lka}
  M.~Duerr and P.~Fileviez P\'erez,
  ``Baryonic Dark Matter,''
  Phys.\ Lett.\ B {\bf 732} (2014) 101
  [arXiv:1309.3970 [hep-ph]].
  
\bibitem{Ohmer:2015lxa}
  S.~Ohmer and H.~H.~Patel,
  ``Leptobaryons as Majorana Dark Matter,''
  Phys.\ Rev.\ D {\bf 92} (2015) no.5,  055020
  [arXiv:1506.00954 [hep-ph]].

\bibitem{Duerr:2016tmh}
  M.~Duerr, F.~Kahlhoefer, K.~Schmidt-Hoberg, T.~Schwetz and S.~Vogl,
  ``How to save the WIMP: global analysis of a dark matter model with two s-channel mediators,''
  JHEP {\bf 1609} (2016) 042
  [arXiv:1606.07609 [hep-ph]].
  
\bibitem{Duerr:2014wra}
  M.~Duerr and P.~Fileviez Perez,
  ``Theory for Baryon Number and Dark Matter at the LHC,''
  Phys.\ Rev.\ D {\bf 91} (2015) no.9,  095001
  [arXiv:1409.8165 [hep-ph]].
  
\bibitem{Ellis:2018xal}
  J.~Ellis, M.~Fairbairn and P.~Tunney,
  ``Phenomenological Constraints on Anomaly-Free Dark Matter Models,''
  arXiv:1807.02503 [hep-ph].
  
\bibitem{ElHedri:2018cdm}
  S.~El Hedri and K.~Nordstrom,
  ``Whac-a-constraint with anomaly-free dark matter models,''
  arXiv:1809.02453 [hep-ph].
  
\bibitem{Caron:2018yzp}
  S.~Caron, J.~A.~Casas, J.~Quilis and R.~Ruiz de Austri,
  ``Anomaly-free Dark Matter with Harmless Direct Detection Constraints,''
  arXiv:1807.07921 [hep-ph].
  
\bibitem{Krovi:2018fdr}
  A.~Krovi, I.~Low and Y.~Zhang,
  ``Broadening Dark Matter Searches at the LHC: Mono-X versus Darkonium Channels,''
  [arXiv:1807.07972 [hep-ph]].
  
\bibitem{Gondolo:2011eq}
  P.~Gondolo, P.~Ko and Y.~Omura,
  ``Light dark matter in leptophobic Z' models,''
  Phys.\ Rev.\ D {\bf 85} (2012) 035022
  [arXiv:1106.0885 [hep-ph]].
 
\bibitem{Batell:2014yra}
  B.~Batell, P.~deNiverville, D.~McKeen, M.~Pospelov and A.~Ritz,
  ``Leptophobic Dark Matter at Neutrino Factories,''
  Phys.\ Rev.\ D {\bf 90} (2014) no.11,  115014
  [arXiv:1405.7049 [hep-ph]].
   
\bibitem{Bishara:2018vix}
  F.~Bishara, J.~Brod, B.~Grinstein and J.~Zupan,
  ``Renormalization Group Effects in Dark Matter Interactions,''
  arXiv:1809.03506 [hep-ph].
  
  
\bibitem{Duerr:2017whl}
  M.~Duerr, P.~Fileviez Perez and J.~Smirnov,
  ``Baryonic Higgs at the LHC,''
  JHEP {\bf 1709} (2017) 093
  [arXiv:1704.03811 [hep-ph]].
  
  
 \bibitem{CMS18.8}
The CMS collaboration, ``Search for narrow resonances in dijet final states at $\sqrt s = 8$ TeV with the novel CMS technique of data scouting'', Phys. Rev. Lett. \tb{117} (2016) 031802 [arXiv:1604.08907 [hep-ex]].


 \bibitem{CMS19.7}
The CMS collaboration, ``Search for narrow resonances in the $b$-tagged dijet mass spectrum in proton-proton collisions at $\sqrt s = 8$ TeV'', Phys. Rev. Lett. \tb{120} (2018) 201801 [arXiv:1802.06149 [hep-ex]].

  

  \bibitem{CMS35.9}
The CMS collaboration, ``Search for low mass vector resonances decaying into quark-antiquark pairs in proton-proton collisions at $\sqrt s = 13$ TeV'', JHEP \tb{1801} (2018) 097 [arXiv:1710.00159 [hep-ex]].

 
   \bibitem{CMS27&36}
The CMS collaboration, ``Search for narrow and broad dijet resonances in proton-proton collisions at $\sqrt s = 13$ TeV and constraints on dark matter mediators and other new particles'', CMS-EXO-16-056, CERN-EP-2018-123 (2018) [arXiv:1806.00843 [hep-ex]]. Submitted to JHEP. 



  \bibitem{ATLAS20.3}
The ATLAS collaboration, ``Search for new phenomena in the dijet mass distribution using $pp$ collision data at $\sqrt s = 8$ TeV with the ATLAS detector'', Phys. Rev. D \tb{91} (2015) 052007 [arXiv:1407.1376 [hep-ex]]. 



   \bibitem{ATLAS3.6&29.3}
The ATLAS collaboration, ``Search for low-mass dijet resonances using
trigger-level jets with the ATLAS detector in $pp$ collisions at $\sqrt s = 13$ TeV'', CERN-EP-2018-033 (2018) [arXiv:1804.03496 [hep-ex]]. Submitted to Phys. Rev. Lett..


 
  \bibitem{ATLAS36.1}
The ATLAS collaboration, ``Search for light resonances decaying to boosted quark pairs and produced in association with a photon or a jet in proton-proton collisions at $\sqrt s = 13$ TeV with the ATLAS detector'', CERN-EP-2017-280 (2018) [arXiv:1801.08769 [hep-ex]]. Submitted to Phys. Lett. B.

  \bibitem{ATLAS37.0}
The ATLAS collaboration, ``Search for new phenomena in dijet events using 37 fb$^{-1}$ of $pp$ collision data collected at $\sqrt s = 13$ TeV with the ATLAS detector'', Phys. Rev. D \tb{96} (2018) 052004 [arXiv:1703.09127 [hep-ex]].


\bibitem{UA2}
  J.~Alitti {\it et al.} [UA2 Collaboration],
  ``A Measurement of two jet decays of the $W$ and $Z$ bosons at the CERN $\bar{p} p$ collider,''
  Z.\ Phys.\ C {\bf 49} (1991) 17.
  
  
\bibitem{CDF}
  F.~Abe {\it et al.} [CDF Collaboration],
  ``Search for new particles decaying to dijets at CDF,''
  Phys.\ Rev.\ D {\bf 55} (1997) R5263
  [hep-ex/9702004].
  
\bibitem{Gondolo:1990dk}
  P.~Gondolo and G.~Gelmini,
  ``Cosmic abundances of stable particles: Improved analysis,''
  Nucl.\ Phys.\ B {\bf 360} (1991) 145.
  
\bibitem{Ade:2013zuv}
  P.~A.~R.~Ade {\it et al.} [Planck Collaboration],
  ``Planck 2013 results. XVI. Cosmological parameters,''
  Astron.\ Astrophys.\  {\bf 571} (2014) A16
  [arXiv:1303.5076 [astro-ph.CO]].
 
\bibitem{Hoferichter:2017olk}
  M.~Hoferichter, P.~Klos, J.~Menendez and A.~Schwenk,
  ``Improved limits for Higgs-portal dark matter from LHC searches,''
  Phys.\ Rev.\ Lett.\  {\bf 119} (2017) no.18,  181803
  [arXiv:1708.02245 [hep-ph]].
  

\bibitem{XENON1T}
  E.~Aprile {\it et al.} [XENON Collaboration],
  ``Physics reach of the XENON1T dark matter experiment,''
  JCAP {\bf 1604} (2016) no.04,  027
  [arXiv:1512.07501 [physics.ins-det]].
  
  
\bibitem{XENONnT}
  E.~Aprile {\it et al.} [XENON Collaboration],
  ``Physics reach of the XENON1T dark matter experiment,''
  JCAP {\bf 1604} (2016) no.04,  027
  [arXiv:1512.07501 [physics.ins-det]].
  
\bibitem{FermiLAT}
  M.~Ackermann {\it et al.} [Fermi-LAT Collaboration],
  ``Searching for Dark Matter Annihilation from Milky Way Dwarf Spheroidal Galaxies with Six Years of Fermi Large Area Telescope Data,''
  Phys.\ Rev.\ Lett.\  {\bf 115} (2015) no.23,  231301
  [arXiv:1503.02641 [astro-ph.HE]].

\bibitem{Perez:2013tea}
  P.~Fileviez Perez and H.~H.~Patel,
  ``Baryon Asymmetry, Dark Matter and Local Baryon Number,''
  Phys.\ Lett.\ B {\bf 731} (2014) 232
  [arXiv:1311.6472 [hep-ph]].

\bibitem{KamionkowskiGriest:1990}
  K.~Griest and M.~Kamionkowski,
  ``Unitarity Limits on the Mass and Radius of Dark-Matter Particles,''
  Phys.\ Rev.\ Lett. {\bf 64} (1990) 6.


\end{thebibliography}
\end{document}